\documentclass[12pt]{article}
\pdfoutput=1
\usepackage{bm}
\usepackage{amsmath}
\usepackage{amsfonts}
\usepackage{amssymb}
\usepackage{comment}
\usepackage{graphicx}
\usepackage{bm}
\usepackage{mathtools}
\usepackage{float}
\usepackage{adjustbox}
\usepackage{cite}
\usepackage{bbold}
\usepackage{multirow}
\usepackage{enumerate}
\usepackage{dynkin-diagrams}
\usepackage{ulem}
\usepackage{color}
\usepackage{stackengine,scalerel}
\newcommand\textaltcolon{\ensurestackMath{\stackon[0.5ex]{\circ}{\circ}}}
\newcommand\altcolon{\savestack\Tmp{\raisebox{-.7pt}{$\textaltcolon$}}%
  \dp\Tmpcontent=\dimexpr\dp\Tmpcontent-.7pt\relax%
  \mathrel{\scalerel*{\Tmp}{:}}}

\usepackage[utf8]{inputenc}
\usepackage[hidelinks]{hyperref}
\usepackage{physics}

\begin{document}
\begin{titlepage}
		\begin{flushright}
			TIT/HEP-703 \\
			August, 2024
		\end{flushright}
		\vspace{0.5cm}
		\begin{center}
			{\Large \bf Integrals of motion in conformal field theory with $W$-symmetry and the ODE/IM correspondence
			}
			\lineskip.75em 
			\vskip 2.5cm
			{\large  Katsushi Ito, Mingshuo Zhu }
			\vskip 2.5em
			{\normalsize\it Department of Physics,\\
				Institute of Science Tokyo\\
				Tokyo, 152-8551, Japan}
			\vskip 3.0em
		\end{center}
\begin{abstract}
We study the ODE/IM correspondence between two-dimensional $WA_{r}$/$WD_{r}$-type conformal field theories and the higher-order ordinary differential equations (ODEs) obtained from the affine Toda field theories associated with $A_r^{(1)}/D_r^{(1)}$-type affine Lie algebras. We calculate the period integrals of the WKB solution to the ODE along the Pochhammer contour, where the WKB expansions correspond to the classical conserved currents of the Drinfeld-Sokolov integrable hierarchies. We also compute the integrals of motion for $WA_{r}$($WD_{r}$) algebras on a cylinder. Their eigenvalues on the vacuum state are confirmed to agree with the period integrals up to the sixth order. These results generalize the ODE/IM correspondence to higher-order ODEs and can be used to predict higher-order integrals of motion.
\end{abstract}
\end{titlepage}

	\baselineskip=0.7cm
    \numberwithin{equation}{section}
	\numberwithin{figure}{section}
	\numberwithin{table}{section}

\section{Introduction}
In two-dimensional conformal field theory (CFT), Integrals of motion (IoMs) play an important role in understanding the dynamics and symmetries of the system\cite{Bazhanov:1994ft}.  There is an infinite number of IoMs due to integrability; hence, their construction is a fundamental problem in CFT. In the minimal model of CFT \cite{Belavin:1984vu}, an explicit form of the IoMs and their eigenvalues in the vacuum state is obtained in terms of the energy-momentum (EM) tensor up to spin eight \cite{Bazhanov:1994ft,Dymarsky:2019iny}. So far, it has been generalized to the $WA_{2}$ model\cite{Fateev:1987vh}, a CFT with a spin 3 field, in
\cite{Kupershmidt:1989bf, Ashok:2024zmw, Bazhanov:2001xm}. 
However, explicit construction of IoMs in higher spin $W\mathfrak{g}$ algebra is still an open question. We also note that IoMs play a crucial role in studying thermal correlators in CFT \cite{Maloney:2018hdg,Dymarsky:2022dhi,Ashok:2024zmw}.

The IoMs in CFT with $W$-symmetry are defined as mutually commuting conserved quantities constructed from the energy-momentum tensor and $W$ currents\cite{Kupershmidt:1989bf}. 
On the complex plane, their commutation relations are obtained from  
the operator product expansions (OPEs) among the currents\cite{Zamolodchikov:1987jf,Sasaki:1987mm,Eguchi:1989hs}.
The computation of the OPE with the $W$-currents can be simplified by the free-field realization \cite{Fateev:1987vh, Fateev1988}. 
Another interesting approach to finding the IoMs in CFT is to use the ODE/IM correspondence, which is the main subject of this paper.

The ODE/IM correspondence \cite{Dorey:1998pt, Bazhanov:1998wj} describes the relation between the spectral analysis of ordinary differential equations (ODEs) and the functional approach of two-dimensional quantum integrable models (IMs). For the Schr\"odinger type second-order ODE, the correspondence shows that the connection coefficients between the asymptotic solutions of the ODE around the singularities define the $Q$-, $T$-, and $Y$- functions of the quantum integrable model, or more precisely, the CFT. 
For example, the Q-function, which connects the basis of the solutions around the origin and infinity, satisfy 
the non-linear integral equations \cite{Bazhanov:1996dr,Dorey:1999uk}.
The zeros of the Q-functions determine the roots of the Bethe ansatz equations.
The Y-functions that satisfy the thermodynamic Bethe ansatz (TBA) equations in the integrable theories \cite{Bazhanov:1996aq} correspond to the Borel resummation of the WKB period of the solutions of the ODE \cite{Ito:2018eon}. Both integral equations provide the effective central charge of the CFT, which is characterized by the lowest conformal dimension of the primary field. 

The IoMs and their eigenvalues on the vacuum state provide much information about CFT with extended symmetry.  The local IoMs are obtained by the T-functions expanded in the power of the spectral parameter. 
For the Schr\"odinger equation with monomial potential, the IoMs of the minimal models \cite{Bazhanov:1994ft} are shown to coincide with higher-order corrections to the specific WKB period for the ODE (see also \cite{Ito:2018eon}).

It is interesting to study a correspondence between the higher-order IoM and the WKB period for higher-order ODE. The ODE/IM correspondence for higher-order ODEs has been studied in \cite{Dorey:1999pv, Suzuki:1999hu, Dorey:2000ma, Dorey:2006an} based on integral equations, which provide a relation to CFT with $W$ currents. 
In particular, the $W\mathfrak{g}$ CFT corresponds to the linear differential system associated with the $\hat{\mathfrak g}^{\vee}$ affine Toda field equations modified by some conformal transformation. Such conformal transformation 
is necessary to produce the appropriate correction to the potential term in the ODE.  
The correspondence between the IoMs in CFT with the $WA_2$ algebra and the WKB periods has been observed in \cite{Bazhanov:2001xm} and \cite{Ashok:2024zmw}. The correspondence for a CFT with the $WA_{r}$ algebra with a specific central charge has also been studied in \cite{Bazhanov:2003ua}.

In the previous paper\cite{Ito:2023zdc}, we studied the WKB expansion of the solutions to the linear differential equations associated with classical affine Lie algebras. 
The linear differential equation can be diagonalized by the gauge transformation, which reduces to a set of first-order linear differential equations. The diagonalized connection in the fundamental representation is a generating function of the conserved currents of the integrable hierarchy in the Drinfeld-Sokolov reduction\cite{Drinfeld:1984qv}. Moreover, the diagonalized connection satisfies the Riccati equation of the 
adjoint ODE, which is also satisfied by the top component of the solution to the adjoint linear problem. The WKB periods can be regarded as the conserved charges of the classical integrable hierarchy, which implies that the ODE/IM correspondence shows a nontrivial relation between classical and quantum theories. 

The purpose of this paper is to find the IoMs in CFT with the $WA_{r}$ or $WD_{r}$ symmetry and explore the correspondence between them and the WKB integrals of the classical conserved charges of the $A_r^{(1)}$ and $D_r^{(1)}$ adjoint linear systems. These two adjoint linear systems are the same as their own ones, whose WKB solution is equivalent to the one for the higher-order ODE satisfied by the top component of the solution of the linear problem.

This paper is organized as follows. In section \ref{Sec: 2}, we first carry out the WKB analysis on the $A_{r}^{(1)}$- and $D_{r}^{(1)}$-type higher-order ODEs given in \cite{Dorey:2006an, Ito:2013aea, Ito:2023zdc}. Both can be obtained from the linear system of the modified affine Toda field equations. We then compute their WKB integrals along the Pochhammer contour up to the 8th order for general rank $r$. In section \ref{Sec: 3}, we first define the CFT on a complex plane and present the EM tensor, $W$ currents, and their OPEs. Then, we introduce the free-field realization in $WA_{r}$ and $WD_{r}$ algebras. Finally, we define the IoMs on a cylinder and compute their eigenvalues on the vacuum state. In section \ref{Sec: 4}, we compare the WKB integrals and the eigenvalue of the IoMs and confirm the ODE/IM correspondence for the linear problem associated with the $A_r^{(1)}$ and $D_r^{(1)}$ -types affine Lie algebras.

\section{Modified affine Toda field equations and the linear system}\label{Sec: 2}
In this section, we first define the linear problem for affine Toda field theories modified by a conformal transformation based on \cite{Lukyanov:2010rn} and \cite{Dorey:2012bx, Ito:2013aea, Adamopoulou:2014fca}. After taking the light-cone and conformal limit, we obtain the higher-order ODE satisfied by the top component of the solution to the linear problem and their WKB solutions. The WKB solutions correspond to the classical conserved currents of the Drinfeld-Sokolov reduction of the soliton hierarchy based on affine Lie algebras \cite{Drinfeld:1984qv}. We then compute the integrals of the WKB solutions along the Pochhammer contour. These classical quantities will be compared with the conserved charges of the quantum IoMs in the CFT with $W$ symmetry, which are computed in the next section.

\subsection{The WKB analysis of linear problem}
Let us begin with the modified affine Toda field equations on the complex plane.
\begin{equation}\label{eq:matfes}
    \partial_{\Bar{z}}\partial_{z}\phi(z,\Bar{z})-\left(\frac{m^2}{\beta}\right)\left[\sum_{i=1}^{r}\alpha_{i}\exp\big(\beta\alpha_{i}\cdot \phi\big)+p(z)\Bar{p}(\Bar{z})\alpha_{0}\exp\big(\beta\alpha_{0}\cdot \phi\big)\right]=0,
\end{equation}
where $\alpha_{i}$($\alpha_{0}$) are simple(affine) roots of affine Lie algebra $\hat{\mathfrak{g}}$. Here the \lq\lq modified" means the appearance of holomorphic functions $p(z)$ and $\Bar{p}(\Bar{z})$ \cite{Lukyanov:2010rn},\cite{Dorey:2012bx, Ito:2013aea}.
The modified affine Toda field equation \eqref{eq:matfes} can be rewritten as the integrability condition for a pair of linear systems 
\begin{align}\label{eq:linsys1}
    {\cal L}\Psi=\bar{\cal L}\Psi=0,
\end{align}
where ${\cal L}$ and $\bar{\cal L}$ denote 
the Lax operators
\begin{align}\label{eq: mdr lax}
\begin{split}
&\mathcal{L}=\partial_{z}+\sum_{i=1}^{r}\beta\partial_{z}\phi_{i}(z)H_{i}+\lambda\left(\sum_{i=1}^{r}E_{\alpha_{i}}+p(z)E_{\alpha_{0}}\right),\\
&\Bar{\mathcal{L}}=\partial_{\Bar{z}}+\lambda^{-1}e^{-\beta\sum_{i=1}^{r}\phi_{i}H_{i}}\left(\sum_{i=1}^{r}E_{-\alpha_{i}}+\Bar{p}(\Bar{z})E_{-\alpha_{0}}\right)e^{\beta\sum_{i=1}^{r}\phi_{i}H_{i}}.
\end{split}
\end{align}
Here $E_{\pm\alpha_{0}}$, $E_{\pm\alpha_{i}}$ and $H_{i}$ ($i=1,\dots, r$) are the Chevalley generators of affine Lie algebra $\hat{\mathfrak{g}}$,
and $\lambda$ is the spectral parameter.
Next, we will take the light-cone limit and the conformal limit of the linear system \eqref{eq:linsys1}.
First, we set 
\begin{equation}
    p(z)=z^{hM}-s^{hM},
\end{equation}
where $h$ is the Coxeter number, $M$ is a positive real number with $M>(1/h-1)$ and $s$ is an arbitrary parameter. The function $p(z)$ determines the behavior of $\phi(z)$ at infinity \cite{Lukyanov:2010rn,Ito:2013aea}, while at the origin, we impose the boundary condition for $\phi(z)$ as \cite{Lukyanov:2010rn}
\begin{gather}\label{eq: phi0}
  \phi(z) = \frac{l}{\beta}\log(z)+\mathcal{O}(1) , 
\end{gather}
with $l$ an $r$-dimensional vector.
We first take the light-cone limit $\bar{z}\rightarrow 0$ and the conformal limit $\lambda\rightarrow \infty$
\begin{equation}\label{eq: con_limit}
    z=\lambda^{-\frac{1}{1+M}}x,\quad s^{hM}=\lambda^{-\frac{hM}{1+M}}E
\end{equation}
with $x$ and $E$ finite. In these limits, the linear problem reduces to a single holomorphic linear differential equation.
Further rescaling $x$ and $E$ as
\begin{equation}\label{eq: rescale}
    x\rightarrow \epsilon^{\frac{1}{1+M}}x,\quad E\rightarrow \epsilon^{\frac{hM}{1+M}}, 
\end{equation}
and making the gauge transformation as \cite{Ito:2020htm}, the holomorphic linear problem \eqref{eq:linsys1} becomes the first-order linear differential equation
\begin{align}\label{eq: linear problem}
\begin{split}
&\epsilon\partial_{x}\Psi+\epsilon\frac{1}{x}(l\cdot H)\Psi+\left(\sum_{i=1}^{r}E_{\alpha_{i}}+p(x)E_{\alpha_{0}}\right)\Psi=0\\
\end{split}
\end{align}
with $p(x)=x^{hM}-1$. 
We will consider the WKB solution of the linear problem \eqref{eq: linear problem} for $\hat{\mathfrak g}=A_r^{(1)}$ and $D_r^{(1)}$.

\subsubsection*{The WKB analysis of the $A_{r}^{(1)}$ type}
Now we focus on the linear problem \eqref{eq: linear problem} in the $(r+1)$-dimensional fundamental representation of $A_r$.
Let $\left\{\epsilon_1, \ldots, \epsilon_{r+1}\right\}$  be  the weights of this representation. They satisfy the conditions:
\begin{align}
\begin{split}
&\epsilon_i \cdot \epsilon_j=\delta_{i j}-\frac{1}{r+1}, \quad \sum_{i=1}^{r+1} \epsilon_i=0,\\
& \sum_{i<j} \epsilon_i \otimes \epsilon_j=-\frac{1}{2} \mathbb{1},\quad \sum_{i=1}^{r+1} i \epsilon_i=-\rho^{\vee}.
\end{split}
\end{align}
The linear equation for $\Psi$ \eqref{eq: linear problem} in the representation reduces to the $(r+1)$-th order ODE for the first component $\psi$ of $\Psi$ \cite{Sun:2012xw,Ito:2023zdc}. 
\begin{equation}\label{eq: Ar ODE}
    \left[\epsilon^{r+1}\left(\partial_{x}-\frac{l_{1}}{x}\right) \left(\partial_{x}-\frac{l_{2}}{x}\right)\dots \left(\partial_{x}-\frac{l_{r}}{x}\right) \left(\partial_{x}-\frac{l_{r+1}}{x}\right)+(-1)^{r}p(x)\right]\psi(x,\epsilon)=0
\end{equation}
with $l_i=\epsilon_i\cdot l$ and $h=r+1$ for $A_{r}^{(1)}$.
This ODE is the same as that appears in the ODE/IM correspondence \cite{Dorey:2006an}. 
To find the WKB solution, it is convenient to expand the first term in \eqref{eq: Ar ODE} and rewrite it in the form 
\begin{equation}\label{eq:expanded_ad_Ar_ode}
\left[(\epsilon\partial_{x})^{h}+\sum_{i=2}^{h}\epsilon^{i}\frac{L_{i}}{x^{i}}(\epsilon\partial_{x})^{h-i}+(-1)^{r}p(x)\right]\psi(x,\epsilon)=0
\end{equation}
with the coefficients $L_i$ some functions of $l$. 
In particular, they are expressed in terms of the 
the elementary symmetric polynomials of $(\epsilon_i, l+\rho)$:
\begin{equation*}\label{eq: momentum l}
    s_{k}=\sum_{1\leq i_{1}<i_{2}<\dots<i_{k}\leq r+1} (\epsilon_{i_{1}}, l+\rho) (\epsilon_{i_{2}}, l+\rho) \dots  (\epsilon_{i_{k}}, l+\rho),
\end{equation*}
where $\rho$ is the Weyl vector of the Lie algebra, the summation of the fundamental weights. Especially for $A_r$ algebras, $\rho=\sum_{i=1}^{r}(r+1-i)\epsilon_i$.

To testify the ODE/IM correspondence for Eq.\eqref{eq:expanded_ad_Ar_ode}, we need an explicit form of $L_i$'s. The first five coefficients are found to be
\begin{align}\label{eq: Ar_L}
    \begin{split}
        L_{2} &=s_{2}+\frac{1}{24}r(r+1)(r+2),\\
        L_{3} &=-s_{3}-(r-1)L_{2},\\
        L_{4} & = s_{4} +\frac{3}{2}(r-2)s_{3} + \frac{1}{24}(r+24)(r-1)(r-2)s_{2}\\
        & + \frac{1}{5760}(5r+228)(r+2)(r+1)r(r-1)(r-2),\\
        L_{5} & = -s_{5}-2(r-3)s_{4}-\frac{1}{24} (r-3) (r-2) (r+44)s_3\\
        & - \frac{1}{12} (r-3) (r-2) (r-1) (r+12)s_{2}\\
        & - \frac{1}{2880}(r-3) (r-2) (r-1) r (r+1) (r+2) (5 r+108),
    \end{split}
\end{align}
and the 6th, 7th, and 8th orders are listed in Appendix \ref{app:S_fun}. 

We now study the WKB solution of \eqref{eq:expanded_ad_Ar_ode}.
Let us consider the WKB expansion of $\psi(x,\epsilon)$,  normalized at some non-zero point on the complex plane\footnote{In the Schr\"odinger equation with centrifugal potential, the WKB approximation requires the potential receives the Kramers-Langer correction to recover the correct boundary condition at the origin\cite{Langer37,langer2}. However, in our analysis, we do not need the WKB solution around the origin and our results do not lead to any modification.}, 
where the parameter $\epsilon$ plays a role of the Planck constant $\hbar$:
\begin{align}\label{eq: WKB expand}
    \psi(x,\epsilon)&=\exp(\frac{1}{\epsilon}\int^x dx\,S(x,\epsilon)), \\ S(x,\epsilon)&=\sum_{i=0}^{\infty}\epsilon^{i}S_{i}(x). \label{eq: WKB expand2}
\end{align}
Substituting Eq.\eqref{eq: WKB expand} into  Eq.\eqref{eq:expanded_ad_Ar_ode}, one obtains the Riccati equation for $S(x,\epsilon)$:
\begin{equation}\label{eq:riccati1}
(\epsilon\partial_{x}+S)^{r}S+\sum_{i=2}^{r+1}\frac{L_{i}}{x^{i}}(\epsilon\partial_{x}+S)^{r-i}S+(-1)^{r}p(x)=0,
\end{equation}
where we define $(\epsilon\partial_{x}+S)^{-1}S=1$ in the summation. Then we expand $S(x,\epsilon)$ as Eq.\eqref{eq: WKB expand2} and solve the equation \eqref{eq:riccati1} recursively. Here the first two terms of order $\epsilon^n$ $(n=0,1)$ are 
\begin{align}
    \begin{split}
        \epsilon^{0}:\;\;(-1)^r p(x)+S_{0}^{r+1} =0,\quad \epsilon^{1}:\;\;\frac{1}{2} (r+1) S_{0}^{r-1} \left(r S_{0}'+2 S_{0}S_1\right)=0.
    \end{split}
\end{align}
The higher-order terms can be determined similarly. Their solutions are given by
\begin{align}\label{Eq. A_r S_fun}
\begin{split}
        S_{0}(x)&=-p(x)^{\frac{1}{h}},\\
        S_{1}(x)&=\frac{h-1 }{2 h}\partial_{x}\log(p),\\
        S_{2}(x)&=\frac{L_2}{h}p^{-\frac{1}{h}}x^{-2}-\frac{(h-1)(h+1)M (hM-1)}{12}p^{-\frac{1}{h}-1} x^{h M-2}\\
        \;\;\;\;\;\;\;& +\frac{(h-1)(h+1) (2h+1) M^2}{24}p^{-\frac{1}{h}-2} x^{2h M-2},\\
        \dots
\end{split}
\end{align}
Note that $S_0(x)$ has $(r+1)$ solutions with phases given by $(r+1)$-th root of unity.
Following the convention in \cite{Ito:2023zdc}, we choose the $-$ sign.
It has been shown that all the $(1+hk)$-th terms ($k\geq 0$) in $S_{i}$ becomes total derivatives \cite{Ito:2021boh, Ito:2023zdc} and hence the corresponding WKB period integral vanishes.
The WKB series of these ordinary differential equations become the classical conserved currents in $A_{r}^{(1)}$-type affine Toda field theories \cite{Ito:2023zdc}. 

\begin{figure}[h]
    \centering
    \includegraphics[width=8cm]{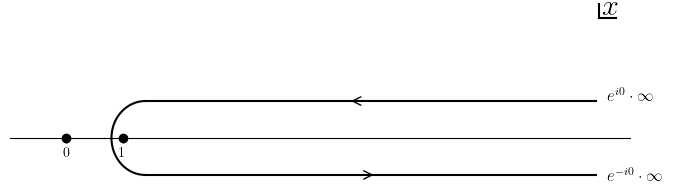}
    \caption{Integral contour $C$}
    \label{fig:enter-label}
\end{figure}

Next, we consider the period integral of the coefficient $S_i(x)$ in the WKB expansion along the cycle $C$. 
Here we choose $C$ as  the Pochhammer contour which starts from $\infty\cdot e^{+i0}$ to $x=1$, goes around $x=1$ and finally ends with $\infty\cdot e^{-i0}$ as in Fig.\ref{fig:enter-label}. 
We define the period integral of $S_i(x)$ as
\begin{equation}\label{eq: Q_def}
    Q_{i} = \oint_{\mathcal{C}} dx\; S_{i}(x).
\end{equation}
For $r=1$, $Q_i$ corresponds to the holomorphic part of the highest-weight quantum conserved charges for the quantum Sinh-Gordon model \cite{Lukyanov:2010rn}.

Now we evaluate the integral \eqref{eq: Q_def} for $A_{r}^{(1)}$ with general $r$. Let us compute the first nontrivial term $Q_{2}(x)$. Since there appear integrants of the form $p(x)^{a}x^{b}$ frequently, it is convenient to introduce the symbol $J(a,b)$:
\begin{equation}\label{eq: recurrence relation}
\begin{aligned}
J(a, b)\equiv\int_C\left(x^{hM}-1\right)^a x^b d x =-\frac{e^{\pi i a} 2 \pi i}{hM} \frac{\Gamma\left(-a-\frac{b+1}{hM}\right)}{\Gamma(-a) \Gamma\left(1-\frac{b+1}{hM}\right)},
\end{aligned}
\end{equation}
where $\Gamma$ is the Gamma function. 
Using the formula $\Gamma(x+1)=x\Gamma(x)$, the integral satisfies the recurrence relation:
\begin{equation}
    J(a+m, b+(hM)n)=e^{i\pi m}\frac{(\frac{b+1}{hM}+n)!(a+m+1)!(a+\frac{b+1}{hM}+1)!}{(a+m+\frac{b+1}{hM}+n+1)!(\frac{b+1}{hM})!(a+1)!}J(a,b,h;M).
\end{equation} 
After evaluating the integral of $S_2(x)$ in Eq.\eqref{Eq. A_r S_fun}, there are three integrals. According to the recurrence relation \eqref{eq: recurrence relation}, they are combined into the following form.
\begin{align*}
    Q_{2} = &\frac{L_{2}}{h}J(-\frac{1}{h},-2)-\frac{(h^2-1)M (h M-1)}{12}J(-\frac{1}{h}-1,-2+hM)\\
    +&\frac{(h^2-1) (2 h+1) M^2}{24}J(-\frac{1}{h}-2,-2+2hM)\\
    =& \frac{1}{h}J(-\frac{1}{h},-2)\Big(L_{2}+\frac{1}{24} (h-1) h(h M-1)\Big).
\end{align*}
This construction rule can be applied to higher orders. For general $Q_{i}(x)$ case, there are $J(a+m,b+(hM)n)$ contained with two integers $0\leq m,n\leq i$. Thanks to the recurrence relation and the construction rule above, it is possible to design an algorithm to compute general $Q_{i}$. Let us list the first five nontrivial terms here, and the $6\text{th}$, $7\text{th}$ and $8\text{th}$ terms in Appendix \ref{app:S_fun}.
\begin{align}\label{eq: Ar_Q}
\begin{split}
        Q_{2}&=\frac{1}{h}J_{1,2}\Big(L_{2}+\frac{1}{24} (h-1) h(h M-1)\Big),\\
        Q_{3}&=\frac{1}{h}J_{2,3}\Big(-L_3-(h-2)L_2 \Big),\\
        Q_{4}&=\frac{1}{h}J_{3,4}\Big(L_{4}+(h-3)\{ -\frac{1}{2h}L_{2}^{2}+\frac{3}{2}L_{3}-\frac{1}{8}[h(M-4)+9]L_{2}\\
        &\quad +\frac{1}{1920}(h-1) [hM-3] [hM-1] [2h(M+1)-1] \}\Big),\\
        Q_{5}&=\frac{1}{h}J_{4,5}\Big(-L_5+\frac{1}{3h}(h-4)\{3L_{2}L_{3}+3(h-2)L_{2}^{2}-6L_{4}\\
        &\quad+[h(M-3)+11]L_{3}+(h-2)(hM+2)L_{2}  \}\Big),
\end{split}
\end{align}
where $L_{k}$ vanishes when $k>h$ and $J(-n/h,m)$ are abbreviated to $J_{n,m}$ for convenience. We can check that $Q_{(1+hk)}$ ($k\in \mathbf{Z}$) becomes zero, which is consistent with the WKB analysis. A similar WKB computation has also been done for $A_{r}^{(1)}$-type ODEs without parameter $M$ in \cite{Bazhanov:2003ua}.

\subsubsection*{The WKB analysis of the $D_{r}^{(1)}$ type}
We want to generalize the WKB analysis for $A_r^{(1)}$-type linear problem to that associated with the $\hat{\mathfrak g}$-type affine Toda field equations has been studied by diagonalization via gauge transformation in \cite{Ito:2023zdc}. This subsection focuses on the $D_r^{(1)}$-type equation, and other types will be studied in a separate paper.  Let us consider the $D_r^{(1)}$-type linear problem with \eqref{eq: mdr lax} for the vector representation. The weight vector of the representation are $\epsilon_1, \dots, \epsilon_r,-\epsilon_r,\dots, -\epsilon_1$, where $\epsilon_i$ are orthonormal basis with  $\epsilon_i \cdot \epsilon_j=\delta_{i j}$. 
The Weyl vector $\rho=\sum_{i=1}^{r}(r-i)\epsilon_i$.

The linear problem reduces to the single differential equation for the component $\psi$ corresponding to the highest weight $\epsilon_1$ \cite{Dorey:2006an}:
\begin{align}\label{eq:ode-dr_limit}
&\epsilon^{2r-2}\left(\partial_{x}-\frac{l_{1}}{x}\right)\cdots \left(\partial_{x}-\frac{l_{r}}{x}\right)\partial_{x}^{-1} \left(\partial_{x}+\frac{l_{r}}{x}\right)\cdots \left(\partial_{x}+\frac{l_{1}}{x}\right)\psi=4\sqrt{p(x)}\partial_{x}\sqrt{p(x)}\psi,
\end{align}
with $l_{k}=\epsilon_{k}\cdot l$ and $h=2r-2$ in $D_{r}^{(1)}$. 

The ODE \eqref{eq:ode-dr_limit} is difficult to solve directly with the WKB method due to the existence of a pseudo-differential operator. However, when we start from the equivalent linear problem \eqref{eq: linear problem},
its WKB solution is found by the diagonalization with gauge transformation \cite{Drinfeld:1984qv}.
Especially for the $D_{r}^{(1)}$-type linear problem in the vector representation, the diagonalized connection is characterized by the two functions: $f(x,\epsilon)$ and $K(x,\epsilon)$ \cite{Ito:2023zdc}. 
The Riccati equations for $f(x,\epsilon)$ and $K(x,\epsilon)$ determine the WKB solution of the linear problem.
Based on this diagonalization method, the WKB solutions to the $D_{3}^{(1)}$ and $D_{4}^{(1)}$ linear problems were found. 
The diagonalization method supports the following prescription to obtain the WKB series $f(x,\epsilon)$ based on an embedding of $B_{r-1}^{(1)}$ into $D_r^{(1)}$. 

First, we set $l_r$ to be zero, by which the pseudo-ODE \eqref{eq:ode-dr_limit} reduces to the $B_{r-1}^{(r)}$-type ODE:
\begin{align}\label{eq:ode_br-1}
  \epsilon^{2r-2}  &\left(\partial_{x}-\frac{l_{1}}{x}\right)\cdots \left(\partial_{x}-\frac{l_{r-1}}{x}\right)\partial_{x} \left(\partial_{x}+\frac{l_{r-1}}{x}\right)\cdots \left(\partial_{x}+\frac{l_{1}}{x}\right)\psi(x)\nonumber\\
  &-4\sqrt{p(x)}\partial_{x}\sqrt{p(x)}\psi(x)=0.
\end{align}
The WKB expansion of this ODE can be studied as in the $A_r^{(1)}$ case.
Expanding the differential operator in \eqref{eq:ode_br-1} such as
\begin{align}
    \left(\partial_{x}-\frac{l_{1}}{x}\right)\cdots \left(\partial_{x}-\frac{l_{r-1}}{x}\right)\partial_{x} \left(\partial_{x}+\frac{l_{r-1}}{x}\right)\cdots \left(\partial_{x}+\frac{l_{1}}{x}\right)
    &=\partial^{2r-1}_{x} + \sum_{k=2}^{2r-1}\frac{L'_{k}}{x^{k}}\partial^{2r-1-k}_{x},
    \label{eq:expbr-1}
\end{align}
and substituting \eqref{eq: WKB expand} into \eqref{eq:ode_br-1}, we
obtain the Riccati equation:
\begin{equation}
    (\epsilon\partial_{x}+S)^{2r-2}S+\sum_{i=2}^{2r-1}\frac{L'_{i}}{x^{i}}(\epsilon\partial_{x}+S)^{2r-2-i}S - 4\sqrt{p(z)}\partial_{z}\sqrt{p(z)} =0.
\end{equation}
The Riccati equation can be solved recursively as
\begin{align}
        S_{0}(x)&=2^{\frac{2}{h}}p(x)^{\frac{1}{h}},\nonumber\\
        S_{1}(x)&=-\frac{1}{2}\partial_{x}\log(p),\nonumber\\
        S_{2}(x)&=-4^{-1/h} \frac{L'_2}{h}p^{-\frac{1}{h}}x^{-2}-2^{-\frac{2}{h}-3}\frac{(h+1) (h+2) (2 h+1)M^{2}}{3}p^{-\frac{1}{h}-2}x^{-2+2hM}\nonumber\\
        \;\;\;\;\;\;\;& +\frac{4^{-\frac{h+1}{h}} (h+1) (h+2)(hM-1)M}{3 h}p^{-\frac{1}{h}-1}x^{-2+hM},
        \label{eq:wkb_br-1}
\end{align}
where $h=2r-2$.
The parameters $L'_k$ in \eqref{eq:expbr-1} are shown to be symmetric polynomials of
$(\epsilon_1,l+\rho)^2,\dots, (\epsilon_{r-1},l+\rho)^2$.
It is observed that the WKB expansion $f(x,\epsilon)$ for $D_r^{(1)}$ can be obtained by promoting $L_k'$ to the symmetric polynomial of the same order by adding $(\epsilon_r,l+\rho)^2$.
Explicitly, the first five $L'_{k}$ for $D_r^{(1)}$ are given by
\begin{align}\label{eq: Dr_L}
\begin{split}
    L'_{2} =&-s_{1}+\frac{1}{12}(2r-1)r(2r-2),\\
    L'_{3} =& -(2r-3)L'_{2},\\
    L'_{4}=  & s_{2} - \frac{1}{6}(r-2) (r+11) (2 r-3)s_{1}\\
    +&\frac{1}{360} (r-2) (r-1) r (2 r-3) (2 r-1) (5 r+109),\\
    L'_{5}= & (10-4 r)L'_4 + 2(r-2) (2 r-5) (2 r-3)L'_2, \\
    L'_{6}= & -s_{3}+\frac{1}{6} (r-3) (r+34) (2 r-5)s_{2}\\
    -&\frac{1}{360} (r-3)(r-2) (2 r-5)(2r-3) (5 r^{2}+339r+1096)s_{1}\\
    +&\frac{1}{45360}(r-3)(r-2) (r-1) r (2 r-1)(2 r-3) (2r-5) (35 r^2+3549 r+22306),
\end{split}
\end{align}
and $s_{k}$ is defined by
\begin{equation*}\label{eq: momentum l D_{r}}
    s_{k}=\sum_{1\leq i_{1}<i_{2}<\dots<i_{k}\leq r} (\epsilon_{i_{1}}, l+\rho)^{2} (\epsilon_{i_{2}}, l+\rho)^{2} \dots  (\epsilon_{i_{k}}, l+\rho)^{2}.
\end{equation*}

The diagonal element $K(x,\epsilon)$ provides another WKB solution whose leading term is given by 
$2p^{-\frac{1}{2}}x^{-r}K_{r}$ where
\begin{equation}
    K_{r} := x^{r}\left(\partial_{x}-\frac{l_1}{x} \right)\cdots \left(\partial_{x}-\frac{l_r}{x} \right)\cdot 1 = (\epsilon_{1}, l+\rho)\cdots (\epsilon_{r}, l+\rho).
\end{equation}
Note that $s_1,\dots,s_{r-1},K_r$ characterize the Casimir invariants of $D_r$.
After substituting these diagonal elements into Eq.\eqref{eq: Q_def}, we can obtain the first five ($Q_{6}$ can be found in the Appendix \ref{app:S_fun}) and the extra WKB integrals:
\begin{align}\label{eq: Dr_Q}
    \begin{split}
        Q_{2} &= - \frac{2^{-\frac{2}{h}}}{h}J_{1,2}\left(L'_{2}+\frac{1}{24} h (h+2) (h M-1)\right),\\
        Q_{3} & = - \frac{2^{-\frac{4}{h}}}{h}J_{2,3}  \left[L'_{3}+(2r-3)L'_{2} \right]= 0,\\
        Q_{4} & = -\frac{2^{-\frac{6}{h}}}{h}J_{3,4}\bigg[L'_4 - \frac{h-3}{2h}L_2^{'2} +\frac{3(h-2)}{2} L'_3 \\
        & - \frac{1}{8} \left[  (M-4)h^2-(6 M-9)h-6\right]L'_2 \\
        &+\frac{1}{1920} h(h-6) (h+2) (Mh-3) (Mh-1) (2(M+1)h-1) \bigg] ,\\
        Q_{5} & = -\frac{2^{-\frac{8}{h}}}{h}J_{4,5}\bigg[ L'_{5}+\frac{4}{h}L'_{2}L'_{3}+2 (h-3)L'_{4} -\frac{(h-4) (h-1)}{h} L_{2}^{'2}\\
        & +\frac{1}{6}\left(40 - 28 h + 6 h^2+(11 - 2 h) hM\right)L'_{3}\\
        & -\frac{1}{6} (h-1) \left((2 h-11)hM-2 (h+2)\right)L'_{2}\bigg],\\
        Q'_{r} & = 2 J(-\frac{1}{2}, -r) (\epsilon_{1}, l_{1}+\rho)\cdots (\epsilon_{1}, l_{r}+\rho), \\
    \end{split}
\end{align}
where $L'_{k}$ also vanishes when $k>h$.

Above all, we have obtained the period integrals of the WKB solutions or charges of the classical conserved currents in modified affine Toda field theories. In the next section, we compute the quantum IoMs in CFT with $W$-symmetry. 

\section{Conformal field theory with $W$ symmetry}\label{Sec: 3}
In this section, we will review two-dimensional CFT with $W$-symmetry and its free field representation. 
For more details about the $W$ algebra, see reviews \cite{Bouwknegt:1992wg, Lukyanov:1990tf}.
We then compute the quantum IoMs on a cylinder by conformal transformation from the ones on a complex plane. We finally compute the zero modes of the IoMs using the normal ordered products of the operators on the complex plane.

Let $(z,\bar{z})$ be a complex coordinate on the complex plane.  A CFT is completely characterized by a set of operators and their operator product expansions (OPE). An operator $A(z,\bar{z})$ in CFT is decomposed into the holomorphic part $A(z)$ and the anti-holomorphic part $\bar{A}(\bar{z})$. We focus on the holomorphic part.

Conformal transformation is generated by the energy-momentum tensor $T(z)$, whose OPE is given by
\begin{equation}
    T(z)T(w)= \frac{c/2}{(z-w)^4}+\frac{2T(w)}{(z-w)^2}+\frac{\partial_{w}T(w)}{z-w}+\cdots,
    \label{eq:opett1}
\end{equation}
where $c$ is the central charge. The term $\dots $ denotes the regular terms.
The primary field $\Phi(z)$ with conformal dimension $\Delta$ is defined by the OPE
\begin{equation}
       T(z)\Phi(w)= \frac{\Delta \Phi(w)}{(z-w)^2}+\frac{\partial_{w}\Phi_(w)}{z-w}+\cdots,
\end{equation}
The symmetry algebra of a CFT can be extended by adding currents. In particular, the algebra generated by higher spin currents is called the $W$ algebra. Let $W_s(z)$ denote the spin $s$ current.  Note that $T(z)$ is a spin 2 current: $W_2(z)=T(z)$.

Let $A(z)$ and $B(z)$ be two symmetry generators with conformal dimension $\Delta_A$ and $\Delta_B$, respectively.
The operator product expansion takes the form
\begin{equation}
    A(z) B(w)=\sum_{1\leq k \leq \Delta_A+\Delta_b} \frac{\{A B\}_k(w)}{(z-w)^k}+\altcolon A B \altcolon (w)+O(z-w).
\end{equation}
Here $\altcolon AB \altcolon(w)$ is the normal ordered product on the complex plane defined by the radial ordered product:
\begin{align}\label{eq:nopr1}
    \altcolon AB \altcolon(w) ={1\over 2\pi i}\oint dz {R(A(z)B(w))\over z-w}. 
\end{align}
 
\subsection{Free field realization of W-algebras }\label{sect:3-1}
We will define the $W$-algebra by the free field realizations. For $A_r$ and $D_r$ type $W$-algebras, their free field realization can be obtained by the quantum Miura transformation \cite{Fateev1988, Lukyanov:1989gg}. The EM tensor and higher spin $W$ currents are obtained from polynomials of the scalar field $\phi(z)$ and their derivatives. The OPE of $\phi_i(z)$ is defined by
\begin{equation}\label{eq: phiphiOPE}
    \phi_i(z) \phi_j(w)=-\delta_{i j} \log (z-w)+\cdots.
\end{equation}
The mode expansion of $\partial \phi_j(z)$
\begin{align}
   i \partial \phi_j(z)&=\sum_{n\in {\mathbb Z}}a_n^j z^{-n-1},
\end{align}
leads to the commutation relations of $a^j_n$: $[a^i_n,a^j_m]=n\delta_{n+m,0}\delta^{ij}$.
The EM tensor in the $W{\mathfrak g}$-algebra associated with a simply-laced Lie algebra ${\mathfrak g}$ is defined by
\begin{align}
    T(z)&=-{1\over2} \altcolon (\partial_{z} \phi)^2(z)\altcolon-i\alpha_0 \rho\cdot \partial^2_{z}\phi(z).
\end{align}
The central charge is given by
\begin{align}
   c&=r-12\alpha_0^2 \rho^2,
\end{align}
where $\rho$ is the Weyl-vector of Lie algebra ${\mathfrak g}$. We also define a vertex operator
\begin{align}
V_{\Lambda}(z)=\altcolon e^{i\Lambda \cdot \phi(z)}\altcolon\;,
\end{align}
which is a primary field with conformal weight $\Delta_\Lambda={1\over2}\Lambda(\Lambda+2\alpha_0\rho)$.

\subsubsection*{Free field realization of $WA_r$ algebra}
The $WA_r$ algebra is generated by spin $2,\dots, r+1$ currents. The quantum Miura transformation \cite{Fateev1988} is a method to obtain their 
free field realization:
\begin{equation}\label{eq: QMT}
(\alpha_0\partial_z)^{r+1}-\sum_{k=2}^{r+1} \Tilde{W}_k(z)\left(\alpha_0 \partial_{z}\right)^{h-k}=\altcolon\left(\alpha_0 \partial_z-\epsilon_1 \cdot  i\partial_{z} \phi\right) \ldots\left(\alpha_0 \partial_z-\epsilon_{r+1} \cdot  i\partial_{z} \phi\right)\altcolon(z)\;.
\end{equation}
Here $h=r+1$.
The current $\Tilde{W}_{2}(z)$ is the energy-momentum tensor $T(z)$, which leads to the central charge
\begin{align}
    c&=r-r(r+1)(r+2)\alpha_0^2.
\end{align} Their OPE can be computed from the OPE \eqref{eq: phiphiOPE}. Applying the differential operator \eqref{eq: QMT} to the vertex operator $V_\Lambda(z)$,  one obtains the formula for conformal weight of $\Tilde{W}_k(z)$ \cite{Fateev1988}:
\begin{align}
\Tilde{\Delta}_k&=(-1)^{k-1} \sum_{i_1<i_2<\dots<i_k}\prod_{j=1}^{k}\left[p_{i_{j}}+\left((k-j)-\frac{1}{2}(r+2-2i_{j})\right)\alpha_{0}\right],
\end{align}
where the new parameter $p$ is given by a shift of $\Lambda$
\begin{equation}\label{eq: p_Lambda}
  \quad p = \Lambda+\alpha_{0}\rho,\quad p_{i}=(\epsilon_{i}, p).
\end{equation}
The conformal weights $\Tilde{\Delta}_{i}$ can be expressed in terms of a symmetric polynomial
\begin{equation}\label{eq: momentum p ar}
    \sigma_{k}=\sum_{1\leq i_{1}<i_{2}<\dots<i_{k}\leq r+1}p_{i_{1}}p_{i_{2}}\dots p_{i_{k}}.
\end{equation}
The first two are given by
\begin{align}\label{eq: Delta_Miura}
\begin{split}
    \Tilde{\Delta}_{2} =  -\sigma_{2}-\frac{1}{4}\binom{r+2}{3}\alpha_0^2,\quad 
    \Tilde{\Delta}_{3} = \sigma_{3}+(r-1) \alpha_0 \sigma_{2}+\binom {r+2}{4}\alpha_0^3.
\end{split}
\end{align}
Usually, the $W$-currents obtained from the quantum Miura transformation \eqref{eq: QMT} are not primary. But we can reconstruct these fields into primary ones denoted by $W_k(z)$. See Appendix \ref{app:pwa} for more detail.

\subsubsection*{Free field realization of $WD_r$ algebra}
The $WD_r$ algebra contains spin $2,4,\dots, 2r-2$ currents $\{W_2,W_4,\dots, W_{2r-2}\}$ and the spin $r$ current $R_r$.
The free field realization of the spin $r$ current is defined by \cite{Lukyanov:1989gg}:
\begin{equation}
R_r(z)=\altcolon\left(\alpha_0 \partial_z-i \epsilon_1 \cdot  \partial_{z} \phi(z)\right) \ldots\left(\alpha_0 \partial_z-i \epsilon_n \cdot  \partial_{z} \phi(z)\right)\altcolon\cdot 1.
\end{equation}
Here $\epsilon_{1},\dots ,\epsilon_r$ are the orthonormal basis of ${\mathbb R}^r$.
The other $W$ currents are determined by the OPE of $R_r$:
\begin{equation}
R_r(z) R_r(w)=\frac{A_r}{(z-w)^{2 r}}+\sum_{k=1}^{r-1} \frac{A_{r-k}}{(z-w)^{2(r-k)}}\left(\Tilde{W}_{2 k}(z)+\Tilde{W}_{2 k}(w)\right),
\label{eq:opev1}
\end{equation}
where
\begin{align}
A_k=\prod_{j=1}^{k-1}(1-2j(2j+1)\alpha_0^2).
\end{align}
The energy-momentum tensor $T(z)$ is given by $\Tilde{W}_2(z)$:
\begin{align}
\Tilde{W}_2(z)&=-{1\over2}\altcolon (\partial_{z}\phi\cdot \partial_{z} \phi)\altcolon(z)-i \alpha_{0} \rho\cdot \partial_{z}^2\phi(z)
\end{align}
with $\rho=\sum_{i=1}^{r}(r-i)\epsilon_i$.
The central charge is 
\begin{align}
    c&=r-r(2r-1)(2r-2)\alpha_0^2.
\end{align}
The free field realization of $\Tilde{W}_{2k}$ can also be obtained from \eqref{eq:opev1}, which are non-primary, while $R_r$ is a primary field.
Here we will construct $\Tilde{W}_4$ and $\Tilde{W}_6$ recursively as follows. First, we note that  $WD_r$ CFT is obtained by adding a free boson to $WD_{r-1}$:
\begin{align}
    WD_r=\{\tilde{\phi}_r \} \oplus WD_{r-1}
\end{align}
with $\tilde{\phi}_i=\epsilon_{r+1-i}\cdot\phi$. Let $W_2^{(r)},\dots, W^{(r)}_{2r-2}, R^{(r)}_r$ be generators of the $WD_r$ algebra. Then $R_r^{(r)}$ is related to $R^{(r-1)}_{r-1}$ via
\begin{align}\label{eq:3-29}
    R^{(r)}_r =-\altcolon\tilde{p}_r R^{(r-1)}_{r-1}\altcolon+\alpha_0 \partial R_{r-1}^{(r-1)},\quad \tilde{p}_r=-i\partial \tilde{\phi}_r.
\end{align}
This implies that the OPE $R_r^{(r)}(z)R_r^{(r)}(w)$ is obtained from $R_{r-1}^{(r-1)}(z)R^{(r-1)}_{r-1}(w)$, which leads to the recursion relations for $W_{2k}^{(r)}$. An explicit calculation and further primary-field construction are shown in the Appendix \ref{app:wdr}. 
There is no explicit formula for conformal weight $\Delta_{i}$ in $WD_{r}$ algebras\cite{Lukyanov:1989gg}, but we can still summarize some lowest-order ones as follows:
\begin{align}
    \Tilde{\Delta}_{2} &
    = {1\over2}\sigma_{1}-{r(2r-2)(2r-1)\over24}\alpha_{0}^2 ,\\
    \Tilde{\Delta}_{4} & 
    = {1\over2}\sigma_{2} -\left({1\over12}(r-7)(r-2)(2r-3)\alpha_{0}^2+{1\over4}\right)\sigma_{1}+{r(2r-1)(r-1)\over24}\alpha_{0}^2 \nonumber\\
&+{1\over720}(r-2)(r-1)r (2r-3)(2r-1)(5r-71)\alpha_{0}^4,
\end{align}
where $\sigma_i$ is defined by
\begin{equation}\label{eq: momentum p}
    \sigma_{k}=\sum_{1\leq i_{1}<i_{2}<\dots<i_{k}\leq r}p_{i_{1}}^{2}p_{i_{2}}^{2}\dots p_{i_{k}}^{2}.
\end{equation}
The conformal weight for the extra generator $R_{r}$ is 
\begin{equation}\label{eq: extra_gen}
    \Delta'_r=p_{1}p_{2}\cdots p_{r}.
\end{equation}

\subsection{Local IoMs and their vacuum eigenvalues on a cylinder}\label{sect:imcyl1}
The IoMs for relevant perturbation of CFT were studied in \cite{ Zamolodchikov:1987jf, Sasaki:1987mm, Eguchi:1989hs,Feigin:1993sb}. The conserved currents that correspond to the generalized KdV flows are constructed
by the OPE between the $W$-currents \cite{Kupershmidt:1989bf}.
Toda field theories, as integrable models, possess an infinite number of IoMs.

On the complex plane, the IoMs can be given by the contour integral around the origin 
\begin{equation}\label{eq: uIoM}
    \mathbf{M}_{k} = \frac{1}{2\pi i}\oint dz\;J_{k}(z).
\end{equation}
Here $J_{k}(z)$ ($k\in {\bf Z}$) are the conserved currents, which can be constructed from normal ordered products of the EM tensor $T$, higher-spin $W$ fields, and their derivatives. 
$J_k(z)$ 
are uniquely determined up to the total derivative terms by the requirement of mutual commutativity condition:
\begin{equation}\label{eq: commutativity relation}
    [\mathbf{M}_{i}, \mathbf{M}_{j}] = 0, \quad \text{for}\; i,j\in \mathbf{Z}.
\end{equation}
The explicit expressions of currents for Virasoro minimal models can be found in \cite{Sasaki:1987mm}.

The IoMs obtained on a complex plane can be transformed into the ones on a cylinder by conformal transformation\cite{Bazhanov:1994ft}. The zero mode of the conserved currents is necessary to see the correspondence between the ODE and the CFT. Due to the difference in the normal ordering prescriptions between the complex plane and cylinder, the zero modes on the cylinder have non-trivial corrections. Some technical details for the Virasoso CFT have been developed in \cite{Dymarsky:2019iny,Novaes:2021vjh}. Here, we will find some formulas of the zero modes for the IoMs in $WA_r$ and $WD_r$ CFTs.  

We consider the conformal transformation defined by
\begin{align}
z=e^u,
\label{eq:confmap1}
\end{align}
where $u$ is the coordinate on a cylinder, and $z$ is the one on a complex plane. Under the conformal transformation, a primary field $A(z)=\sum_n A_n z^{-n-\Delta_A}$ with conformal weight $\Delta_A$ on the complex plane transforms as
\begin{align}
    \hat{A}(u)=\left({dz\over du}\right)^{\Delta_A} A(z)=z^{\Delta_A}
    A(z)=\sum_n A_n z^{-n}.
\end{align}
For a non-primary field, the transformation rule is written as
\begin{align}
    \hat{A}(u)=z^{\Delta_A}
    A(z)+\delta A(z)=:\sum_n \hat{A}_n z^{-n},
    \label{eq:confnp1}
\end{align}
where $\delta A(z)$ is a contribution from the non-primary nature of 
the operator. For the energy-momentum tensor $T(z)$, one finds
\begin{align}
    \hat{T}(u)&=z^2 T(z)-{c\over24},
\end{align}
with
$\delta T(z)=-{c\over 24}$ from the Schwarzian derivative. We define $A_R(z)=A(z)+z^{-\Delta_A}\delta A(z)$ for later convenience.

Under the conformal transformation \eqref{eq:confmap1}, a conserved current $J(z)$ is transformed into $\hat{J}(u)$. The terms in $\hat{J}(u)$ consist of the normal ordered product among the generators of the chiral algebra on the cylinder, where the normal ordered product of operators $\hat{A}$ and $\hat{B}$ is defined by the time-ordered product $T(\hat{A}(u)\hat{B}(v))$ for the time coordinates  ${\rm Re}\;u$ and ${\rm Re}\;v$ on the cylinder.
Then the normal ordered product  $:\hat{A}\hat{B}:(v)$ on the cylinder is defined by
\begin{align}
:\hat{A}\hat{B}:(v)&={1\over 2\pi i}\oint_v du {T(\hat{A}(u)\hat{B}(v))\over u-v},
\end{align}
which differs from the normal ordered product \eqref{eq:nopr1} based on the radial ordering on the complex plane.
Under the conformal transformation \eqref{eq:confmap1} and \eqref{eq:confnp1}, the normal ordered product on the cylinder is expressed as
\begin{align}
    :\hat{A}\hat{B}:(v)&={1\over 2\pi i}\oint_w {dz\over z}
    {z^{\Delta_A}w^{\Delta_B}R(A_R(z)B_R(w))\over \log{z \over w}}.
    \label{eq:opeab1}
\end{align}
The normal ordering \eqref{eq:opeab1} can be written in terms of the normal ordered products on the complex plane, which can be calculated even for higher-spin cases with programming. As we will see later, this reformulation allows us to express the vacuum eigenvalues of the local IoMs in terms of conformal weights. Especially, we found a new formula for the normal ordering $:\hat{W}_3\hat{W}_3:(v)$ with this technique. See Appendix \ref{app:ope1} for more detail.

Based on the normal ordering \eqref{eq:opeab1} and the technique in Appendix \ref{app:ope1}, we can obtain the conserved currents on the cylinder which consist of the EM tensor $\hat{T}(v)$, higher-spin currents $\hat{W}_k(v)$ and their normal orderings. 
Substituting the mode expansions of the W-currents, we find the mode expansion of the conserved currents and finally the IoMs on the cylinder after contour integral. Details are explained in Appendix \ref{app:ope1}. 

\paragraph{Integrals of motion in $WA_r$ CFT}
Let us first begin with the conserved currents $\hat{J}_k(v)$ $(k=2,3,\dots$) in the CFT with $WA_r$-algebra symmetry, where $\hat{J}_{[1+(r+1)i]}(v)$ does not exist for $i\in \mathbf{Z}$  . From the OPE analysis, the first four currents are found to be
\begin{align}
    \hat{J}_2(v)&=\hat{T}(v),\nonumber\\
    \hat{J}_3(v)&=\hat{W}_3(v),\nonumber\\
    \hat{J}_4(v)&=\hat{W}_4+ a_1 :\hat{T}\hat{T}:(v),\nonumber\\
    \hat{J}_5(v)&=\hat{W}_5+b_1: \hat{T}\hat{W}_3:(v),\label{eq:imwar1}
\end{align}
where $a_1$ and $b_1$ can be inferred from low-rank results ($r=4,5,6,7,8$)
\begin{align}\label{eq: Q45_c}
\begin{split}
    a_1&=\frac{3 (r-4) \left[(r+2)c+8 r^2-18 r+4\right]}{2 (5 c+22) (r-2) (r-1) r},\\
     b_1&=\frac{8 (r-5) \left[(r+3)c +15 r^2-33 r+6\right]}{(7 c+114) (r-2) (r-1) r}.
\end{split}
\end{align}
The $6\text{th}$ IoM takes the general form:
\begin{align}
    \hat{J}_6&=\hat{W}_6 +c_1 :\hat{T}\hat{W}_4:+c_2 :\hat{W}_3 \hat{W}_3:
    +c_3 :\hat{T}(:\hat{T}\hat{T}:):+c_4 :\partial \hat{T}\partial \hat{T}:,
\end{align}
where $c_i$'s are some constants.
The computation of the coefficients requires more computational effort.
Here we present examples of rank $r=1,2,3$. For $r=4$, $\hat{J}_6$ does not exist.
\begin{itemize}
\item $r=1$
\begin{align}
    \hat{J}_6&=:\hat{T}(:\hat{T}\hat{T}:):-{c+2 \over12}:\partial \hat{T}\partial \hat{T}:.
\end{align}
\item $r=2$ 
\begin{align}
\hat{J}_6&=:\hat{W}_3 \hat{W}_3:
    +{1\over9} :\hat{T}(:\hat{T}\hat{T}:):+\frac{10-c}{288} :\partial \hat{T}\partial \hat{T}:.
\end{align}
\item $r=3$
\begin{align}
    \hat{J}_6&=:\hat{T}\hat{W}_4:+:\hat{W}_3 \hat{W}_3:
    +\frac{7 c+114}{20 (5 c+22)} :\hat{T}(:\hat{T}\hat{T}:):-\frac{3 c^2+44 c-164}{240 (5 c+22)}:\partial \hat{T}\partial \hat{T}:.
\end{align}
\end{itemize}
We note that the conserved currents on the cylinder can also be directly calculated by the technique of the screening operators. The first three orders have been given in \cite{Lukyanov:1990tf}. The expressions \eqref{eq:imwar1} actually share the same form as the ones on a complex plane.

The conserved charges are defined by the integral of the conserved current on the circle $u=i\sigma$ $(0\leq \sigma\leq 2\pi)$:
\begin{align}
\mathbf{I}_s&=\int_{0}^{2\pi} {d\sigma\over 2\pi}\hat{J}_s(i\sigma).
\end{align}
Substituting the mode expansions $\hat{T}(u)=\sum_n \hat{L}_n e^{-nu}$ and $\hat{W}_s(u)=\sum_n (\hat{W}_s)e^{-nu}$, one finds that $\mathbf{I}_s$ is expressed in terms of the 
modes\footnote{Usually, ${\bf I}_s$ is labelled by the subscript $s-1$ in the literature \cite{Bazhanov:1994ft}.}. 
In particular, when we evaluate 
the eigenvalue of $\mathbf{I}_s$ on the highest weight state $|\{\Delta_s\}\rangle$ corresponding to the primary field $\Phi(z)$ satisfying
\begin{align}
(\hat{W}_s)_0 |\{\Delta_s\}\rangle=\Delta_s |\{\Delta_s\}\rangle.
\end{align}
The eigenvalues depend on the data $\Delta_s$.
Alternatively, we can evaluate
the eigenvalue of ${\bf I}_s$ on the state $|p_i\rangle$ in the momentum basis:
\begin{align}
(\epsilon_i\cdot \hat{a})_0 |p_i\rangle&=p_{i} |p_i\rangle,
\end{align}
where $\hat{a}_{0}$ is the zero mode of $\phi(u)$ based on Eq.\eqref{eq:confnp1}, 
and $p_{i}=\Lambda_{i}+(\epsilon_{i},\alpha_{0}\rho)$ as we have introduced in Eq.\eqref{eq: p_Lambda}.
Next, we evaluate the eigenvalue of the IoMs on the state $|\{\Delta_s\}\rangle$:
\begin{align}
    \mathbf{I}_s |\{\Delta_s\}\rangle&=I_s |\{\Delta_s\}\rangle.
\end{align}
For $I_s$ in \eqref{eq:imwar1}, the eigenvalues are
\begin{align}\label{eq: Delta_i}
    I_2&=\Delta_2-{c\over24},\\
    I_3&=\Delta_3, \label{eq:delta3}\\
    I_4&=\Delta_4+a_1\left(  \Delta_2^2-{c+2\over12}\Delta_2+{5c^2+22c \over 2880}\right),\\
    I_5&=\Delta_5+b_1\left( \Delta_3\Delta_2-{6+c\over 24}\Delta_3\right).
\end{align}
For degree 6, we will show the eigenvalue for $r=1,2,3$, where $r=1,2$ cases have been known in previous literature \cite{Bazhanov:1994ft,Bazhanov:2001xm} :
\begin{itemize}
    \item $r=1$ 
\begin{align}\label{r1_i6}
I_6&=\Delta_{2}^3-{c+4\over 8}\Delta_{2}^2+{(c+2)(3c+20)\over 576}\Delta_{2}-{c(3c+14)(68+7c)\over290304}.
\end{align}
\item $r=2$
\begin{align}\label{r2_i6}
    I_6&=\Delta_3^2
+{1\over 9}\Delta_2^3-{1\over 72}(c+8)\Delta_2^2+{c+2\over 1728}(c+15)\Delta_2
-{c(c+23)\over 870912}(7c+30).
\end{align}
\item $r=3$
\begin{align}\label{r3_i6}
I_6&=\Delta_3^2-{c+12\over12}\Delta_4+2\Delta_4 \Delta_2+{114+7c \over 20(22+5c)}\Delta_2^3
+y_1 \Delta_2^2+y_2 \Delta_2+y_3,
\end{align}
where
\begin{align}
y_1&=-{2264+554c+21c^2 \over 248(22+5c)},\\
y_2&=-{(c+2)(20808+3754c+105c^2) \over 57600(22+5c)},\\
y_3&=-{c(116+3c)(54+7c) \over 4147200}.
\end{align}
\end{itemize}

\paragraph{Integrals of motion in $WD_r$ CFT}
First, we write down the conserved currents in the CFT with $WA_r$-algebra symmetry.
From the OPE analysis in the previous subsection, the conserved currents are summarized as
\begin{align}
    \hat{J}_2(v)&=\hat{T}(v),\nonumber\\
    \hat{J}_4(v)&=\hat{W}_4+ a_1 :\hat{T}\hat{T}:(v),\label{eq:imwdr1}
\end{align}
where $a_1$ can be inferred from low-rank results ($r=4,5,6,7,8$)
\begin{align}\label{eq: D_Q4_c}
     a_1&=-\frac{3 (r-4) \left(2 c r+c+16 r^2-10 r\right)}{2 (5 c+22) (r-1) r (2 r-1)}.
\end{align}
For the spin 6 conserved current in $WD_4$ algebra, it is given by
\begin{align}
    \hat{J}_6&=\hat{W}_6 +x_1 :T(:TT:)+x_2 :\partial T\partial T:,
\end{align}
where
\begin{align}
x_1&=\frac{(11 c+656) (52 c+23)}{2646 (2 c-1) (7 c+68)},\quad x_2=-\frac{(11 c+656) \left(11 c^2-654 c-2432\right)}{42336 (2 c-1) (7 c+68)}.
\end{align}

Next, we evaluate the eigenvalue of the IoMs on the state $|\{\Delta_s\}\rangle$:
\begin{align}
    \mathbf{I}_s |\{\Delta_s\}\rangle&=I_s |\{\Delta_s\}\rangle.
\end{align}
For $\mathbf{I}_s$ in \eqref{eq:imwar1}, the eigenvalues are
\begin{align}\label{eq: D_Delta_i}
    I_2&=\Delta_1-{c\over24},\\
    I_4&=\Delta_4+a_1 \left(  \Delta_2^2-{c+2\over12}\Delta_2+{5c^2+22c\over 2880} \right).
\end{align}
Finally, in $WD_{4}$ algebra, one can obtain
\begin{align}\label{eq: D4_Delta_i6}
I_6&=\Delta_{6}+{(656+11c)(23+52c)\over 2646(2c-1)(68+7c)}\Delta_{2}^3
-{(c+4)(656+11c)(23+52c)\over 21168(2c-1)(68+7c)}\Delta_{2}^2
\nonumber\\
&+{(656+11c)(-96+364c+231c^2+26c^3)\over 254016 (2c-1)(68+7c)}\Delta_{2}
-{c(656+11c)(60+13c)\over 128024064}.
\end{align}

\section{The ODE/IM correspondence}\label{Sec: 4}
In the previous section, we obtained the general form of the vacuum eigenvalues of some IoMs. In this section, we will study the ODE/IM correspondence by comparing the eigenvalues of IoMss in $WA_{r}$($WD_{r}$) algebras with the WKB integral for $A_{r}^{(1)}$($D_{r}^{(1)}$) modified affine Toda field equations. Our result can be viewed as a generalization of \cite{Bazhanov:1994ft, Bazhanov:2001xm, Ashok:2024zmw}, where the correspondence was studied for $WA_{1}$ and $WA_{2}$.
In the previous works \cite{Dorey:2006an, Dorey:2000ma, Ito:2020htm}, the ODE/IM correspondence was studied from the Non-Linear Integral Equations (NLIE) satisfied by the Q-functions and the connection coefficients of the solutions of the ODE. In particular, for the linear problem associated with an affine Lie algebra $\mathfrak{g}$ with potential $p(z)=z^{hM}-E$, the effective central charge 
$c_{\text{eff}}$ is evaluated as
\begin{align}
    c_{\text{eff}}&=r-\frac{12}{h^2 (M+1)}(l +\rho^{\vee})^2.
\end{align}
Comparing with the formula $c_{\text{eff}}=c-24\Delta_0$,
where $c$ is the central charge and $\Delta_0$ is the conformal dimension of the vacuum state, we obtain the relation between the parameters in ODE and CFT:
\begin{align}
    c&=r-12\alpha_0^2\rho^2, \quad \Delta_0=\frac{1}{2} \alpha_0 l \cdot(\alpha_0 l+\alpha_0\rho)
\end{align}
with
\begin{align}\label{eq:alpha0hm}
    \alpha_0^2&={1\over h^2(M+1)}.
\end{align}
The effective central charge is the eigenvalue of the IoM: $-{\bf I}_2/24$. Comparison between the higher IoMs and the integrals of the WKB solutions provides a further non-trivial check for the ODE/IM correspondence. 

\subsection{The ODE/IM correspondence for $A_r^{(1)}$-type ODEs}
Let us first consider the ODE \eqref{eq: Ar ODE}.
The integrals of the WKB expansions $Q_{k}$ ($k=2,3,\dots$) are given in Eq.\eqref{eq: Ar_Q}. We compare these WKB integrals with the vacuum eigenvalue $I_k$ of IoMs ${\bf I}_k$. Since we use the free field representation of $W$ algebra, it is convenient to use momentum basis, where $I_k$ can be expressed in terms of symmetric polynomials in $p$.

\subsubsection*{The WKB integral $Q_{2}$}
Let us consider $Q_{2}$, which is expressed as 
\begin{align}\label{eq: Ar cIoM}
    \begin{split}
        Q_{2} =-h(M+1)J_{1,2}\left[ -\frac{1}{h^2(M+1)}s_{2}-\frac{r}{24} \right].
    \end{split}
\end{align}
Compared with
\begin{equation}
     I_2=-\sigma_{2}-\frac{r}{24},
\end{equation}
it is found that
\begin{equation}
    Q_{2} = h(M+1)J_{1,2}\; I_{2}
\end{equation}
by identifying
\begin{align}
    s_2&=h^2(1+M)\sigma_2,
\end{align}
which implies
\begin{equation}\label{eq: ODE/IM}
    p =\frac{1}{h\sqrt{1+M}}(l+\rho).
\end{equation}
in the momentum basis.
Eq.\eqref{eq: ODE/IM} leads to the general formula
\begin{equation}\label{eq:symm1}
    s_{k} = \left(h\sqrt{1+M}\right)^{k}\sigma_{k}
\end{equation}
for $k\geq 2$.

\subsubsection*{The WKB integral $Q_{3}$}
We will check the formula \eqref{eq:symm1} for higher-order IoMs.
After substituting $L_{3}$ from Eq.\eqref{eq: Ar_L} into Eq.\eqref{eq: Ar_Q}, the WKB integral $Q_{3}$ is given by
\begin{align}
\begin{split}
    Q_{3} = \frac{1}{h}J_{2,3}s_{3}.
\end{split}
\end{align}
Then from Eq.\eqref{eq:symm1} and $I_3=\sigma_3$ from Eq.\eqref{eq: Delta_i}, we find
\begin{equation}
   Q_{3} = h^{2}(1+M)^{\frac{3}{2}}J_{2,3}\; I_{3}.
\end{equation}
Finally, particular attention should be given that $s_{k}$ also vanishes when $k>h$. This condition is also applied to $Q_{k}$ ($k>3$) unless an explicit statement. 

\subsubsection*{The WKB integral $Q_{4}$}
The comparison of the fourth-order IoM is crucial since it includes a constant term. This relation leads to a new equation for the parameters.
We note that the formula for $I_{4}$ and $I_5$ in Eq.\eqref{eq: Delta_i} is only inferred from the low-rank computation ($r\leq 9$). So the correspondence predicts the IoMs for higher-rank 
$WA_r$ algebra with $r\geq 10$. 
According to Eq.\eqref{eq: Ar_L} and Eq.\eqref{eq: Ar_Q}, we rewrite $Q_{4}$ in terms of $s_{i}$ as
\begin{align}
\begin{split}
    Q_{4}=J_{3,4}\left[\frac{1}{h}s_{4} -(h-3)\left(\frac{1}{2h^{2}}s_{2}^2 + \frac{1+M}{8h}s_{2}-\frac{(1+M)^{2}}{1920}h^2(h-1)(\frac{2hM^2}{1+M} -9)  \right)\right].
\end{split}
\end{align}
While $I_4$ in Eq.\eqref{eq: Delta_i} in terms of $\sigma_i$ becomes
\begin{equation*}
    I_4=-\sigma_4+{r-2 \over 2(r+1)}\sigma_2^2+{r-2 \over 8(r+1)}\sigma_2+{(r-2)r(9-2(r+1)\alpha_0^2) \over 1920 (1+r)}.
\end{equation*}
After applying Eq.\eqref{eq:symm1}, we can see that the coefficients for $\sigma_{2}^2$ and $\sigma_{2}$ have already matched. The correspondence
\begin{equation}
    Q_{4}=h^3(1+M)^{2}J_{3,4}\; I_{4}
\end{equation} 
is satisfied when
\begin{equation}
    {1\over 1920h}(h-3)(h-1)(9-2h\alpha_0^2) = \frac{1}{1920h}(h-3)(h-1)\left(9-\frac{2hM^2}{1+M}\right),
\end{equation}
which is the same as the relation \eqref{eq:alpha0hm}. 
Because $\alpha_{0}=\beta-\beta^{-1}$, it is expressed as
\begin{equation}\label{eq: M_beta}
    M = \beta^{-2}-1 \; \text{or}\; M= \beta^2 -1.
\end{equation}
We will choose the first equality in the remaining part because of its appearance in \cite{Ito:2020htm}, where the correspondence is shown from the Bethe ansatz equations.

Finally, based on the relations \eqref{eq: Delta_Miura}, \eqref{eq: Delta_Primary_Miura} between $\Delta_{i}$ and $\sigma_{i}$, one can check the coefficient $a_{1}$ in Eq.\eqref{eq:imwar1} agrees with that we inferred in Eq.\eqref{eq: Q45_c}.

\subsubsection*{The WKB integral $Q_{5}$}
Next, we will check the correspondence for the fifth-order IoMs.
Substituting Eqs.\eqref{eq: ODE/IM} and \eqref{eq: M_beta} into $Q_{5}$ \eqref{eq: Ar_Q}, one can obtain
    \begin{align}
    \begin{split}
        Q_{5}= J_{4,5}\left(\frac{1}{h}s_{5}-\frac{h-4}{h^2}s_2 s_3+\frac{1}{3}(h-4)(1+M)s_{3}\right).
    \end{split}
    \end{align}
From Eq.\eqref{eq: ODE/IM}, one obtains
    \begin{align}
    \begin{split}
        Q_{5}= h^4\beta^{-5}J_{4,5}\; I_{5},
    \end{split}
    \end{align}
where $I_{5}$ in Eq.\eqref{eq: Delta_i} is given by
\begin{equation}
    I_5=\sigma_5-{r-3\over r+1}\sigma_2 \sigma_3-{r-3\over 3(r+1)}\sigma_3.
\end{equation} 
This relation also implies that the coefficient $b_{1}$ in Eq.\eqref{eq:imwar1} is valid for general $r$.

\subsubsection*{The WKB integral $Q_{6}$}
Unlike the order $k\leq 5$  IoMs, we have only the formula of $i_6$ for $r\leq 3$ and do not find a general expression for general rank. 
On the other hand, for general rank $r$, the WKB integral $Q_{6}$ in terms of $s_{i}$ is given by
\begin{align*}
    Q_{6}=J_{5,6}&\bigg[\frac{1}{h}s_{6}+(h-5)\bigg(\frac{1}{h}s_{2}s_{4}+\frac{1}{2h}s_{3}^{2}+\frac{2h-5}{6h^2}s_{2}^{3}-\frac{5}{8}h(1+M)s_{4}\\
    +&\frac{5}{48}(3 h-7)(1+M)s_{2}^{2}-h^{3}(1+M)^{2}c_{6}^{(1)}s_{2}+h^{5}(1+M)^{3}c_{6}^{(2)}\bigg)\bigg]
\end{align*}
with the coefficients $c_{6}^{(1)}$ and $c_{6}^{(2)}$
\begin{align*}
    &c_{6}^{(1)}=\frac{1}{1152 h}[2 h (5 h-11)\frac{M^2}{(1+M)^2}-91h+205],\\
    &c_{6}^{(2)}=\frac{1}{580608h}(h-1)[8 h^2 (h^2+6h-19)\frac{M^4}{(1+M)^4}-5(61 h-139)(2 h\frac{M^2}{(1+M)^2}-5) ].
\end{align*}
After substituting the equation \eqref{eq: ODE/IM} and $ M = \beta^{-2}-1$, it becomes
\begin{align}
\begin{split}
    Q_{6}=-\frac{h^5}{\beta^6}J_{5,6}\bigg[-\sigma_{6}-\frac{h-5}{h}\bigg(&\sigma_{2}\sigma_{4}+\frac{1}{2}\sigma_{3}^{2}+\frac{2h-5}{6h}\sigma_{2}^{3}-\frac{5}{8}\sigma_{4}\\
    &+\frac{5(3 h-7)}{48 h}\sigma_{2}^{2}-c_{6}^{(1)}\sigma_{2}+c_{6}^{(2)}\bigg)\bigg].
\end{split}
\end{align}
When $r=2$ or $3$, where $I_6$ is given by \eqref{r1_i6} and \eqref{r2_i6}, we find the relation
\begin{equation*}
    Q_{6}=-h^5\beta^{-6}J_{5,6}\;I_{6}
\end{equation*}
holds.
Therefore, we conjecture that the vacuum eigenvalue of the IoM ${\bf I}_6$ is 
\begin{equation}\label{eq: Ar_i6}
    I_{6}=-\sigma_{6}-\frac{h-5}{h}\bigg(\sigma_{2}\sigma_{4}+\frac{1}{2}\sigma_{3}^{2}+\frac{2h-5}{6h}\sigma_{2}^{3}-\frac{5}{8}\sigma_{4}
    +\frac{5(3 h-7)}{48 h}\sigma_{2}^{2}-c_{6}^{(1)}\sigma_{2}+c_{6}^{(2)}\bigg)\bigg]
\end{equation}
for general $r$.

Based on the above observations, the ODE/IM correspondence between the WKB integrals and the IoMs is expressed as 
\begin{equation}
    Q_{k}= \frac{(-h)^{k-1}}{\beta^{k}}J_{k-1, k}\;I_{k}.
\end{equation}
Finally, we notice that such correspondence in $WA_{r}$ algebra has also been studied up to the fifth order when $\alpha_{0}=0$ \cite{Bazhanov:2003ua}.

\subsection{The ODE/IM correspondence for $D_{r}^{(1)}$-type pseudo ODEs}
We now apply the same analysis to test the correspondence for $WD_r$ algebra and $D_r^{(1)}$-type affine Toda field equation. 
\subsubsection*{The WKB integral $\mathbf{Q}_{2}$}
Let us begin with the WKB integral $Q_{2}$. Inserting Eq.\eqref{eq: Dr_L} into Eq.\eqref{eq: Dr_Q}, we can see 
\begin{equation*}
    Q_{2} = 2^{-\frac{2}{h}+1}J_{1,2}\; \left( {1\over2h}s_{1} -\frac{r}{24}h(1+M) \right).
\end{equation*}
Compared with the IoM from \eqref{eq: D_Delta_i} in terms of $\sigma_{i}$
\begin{equation*}
  I_2 = {1\over2}\sigma_{1}-\frac{r}{24},
\end{equation*}
one can obtain the equality 
\begin{equation}
    Q_{2} = 2^{-\frac{2}{h}+1}h(1+M) J_{1,2}\; I_2.
\end{equation}
if the following condition is satisfied
\begin{equation}\label{eq:symm2}
    s_{k}=\left(h^{2}(1+M)\right)^{k}\sigma_k
\end{equation}
for $k=1$.
The relation \eqref{eq:symm2} is satisfied if the momentum $p$ satisfies 
\begin{equation}\label{eq: D_r_ODE/IM}
    p =\frac{1}{h\sqrt{1+M}}(l+\rho).
\end{equation}
In summary, we obtain the correspondence $ Q_{2} = 2^{-\frac{2}{h}+1}h(1+M)J_{1,2}\; I_2$, and the relation between $M$ and $\beta$ will be determined in $Q_{4}$

\subsubsection*{The WKB integral $\mathbf{Q}_{4}$}
Since there are no third-order IoMs, we proceed to the fourth-order one.
The WKB integral $Q_{4}$ in \eqref{eq: Dr_Q} in terms of momentum becomes
\begin{align*}
    Q_{4}=&  -2^{-\frac{6}{h}}J_{3,4}\\
   \times  &\left[ \frac{1}{h}s_2 -\frac{h-3}{2h^2}s_{1}^{2}+\frac{h-6}{8}(1+M)s_1 + \frac{h(h-6)}{1920}(h+2)(1+M)^2 \left(\frac{2hM^2}{1+M}-9\right) \right],
\end{align*}
and the IoM in Eq.\eqref{eq: D_Delta_i} in terms of $\sigma_{i}$ becomes
\begin{equation}
    I_4={1\over2}\sigma_2-{2r-5 \over 8(r+1)}\sigma_1^2+{r-4\over 16(r-1)}\sigma_1-{r(r-4)(9-4(r-1)\alpha_0^2) \over 1920(r-1)}.
\end{equation}
After applying \eqref{eq:symm2}, one can see the correspondence
\begin{equation*}
    Q_{4}=-2^{-\frac{6}{h}+1}h^3(1+M)^{2}J_{3,4}\; I_{4}
\end{equation*}
is satisfied when 
\begin{equation}
    {1\over 1920h}(h-6)(h+2)(2ha^2-9) = \frac{1}{1920h}(h-6)(h+2)\left(\frac{2hM^2}{1+M}-9\right).
\end{equation}
This equation implies that $\alpha_0$ satisfies Eq.\eqref{eq:alpha0hm}. $M$ is given by Eq.\eqref{eq: M_beta}, where we will choose the first equality as in the case of $WA_{r}$ algebra.

Finally, if we rewrite $Q_{4}$ in terms of the highest-weight eigenvalues of primary $W$ fields, 
\begin{equation}
    Q_{4}= -2^{-\frac{6}{h}+1}\frac{h^{3}}{\beta^4}J_{3,4}\left( \Delta_{4} + c_{1}\left(\Delta_{2}^2 -{c+2\over12}\Delta_{2}+{5c^2+22c\over 2880}\right) \right),
\end{equation}
where the coefficient $c_{1}$ is nothing but the one we inferred in Eq.\eqref{eq: D_Q4_c}.

\subsubsection*{The WKB integral $\mathbf{Q}_{6}$}
Since $Q_5$ is absent in the WKB expansion, we will present the sixth-order result, which is given by
\begin{align}\label{eq: Dr_Q6}
\begin{split}
    Q_{6}=2^{\frac{10}{h}}J_{5,6} \bigg[\frac{1}{h}s_{3}&-\frac{h-5}{h^2}\bigg(s_{1}s_{2}-\frac{2h-5}{6h}s_{1}^{3}+\frac{1}{48}5h(3 h-10)(1+M)s_{1}^{2}\bigg)\\
    &+\frac{5}{8}(h-6)(1+M)s_{2}+h^3(1+M)^{2}c_{6}^{(1)}s_{1}+h^5(1+M)^3 c_{6}^{(2)}\bigg]
\end{split}
\end{align}
with
\begin{align*}
    c_{6}^{(1)}=\frac{1}{1152 h^2}[(-&200 h + 72 h^2 - 10 h^3)\frac{M^2}{(1+M)^2}+91 h^2-780 h+2300],\\
    c_{6}^{(2)}=\frac{1}{580608h^2}&[-8 h^2 ( h^4- 50 h^2+144h + +472)\frac{M^4}{(1+M)^4}\\
    &+(h+2)(61 h^2-564h+1700) (10h \frac{M^2}{(1+M)^2}-25) ].
\end{align*}
After substituting the equation \eqref{eq: D_r_ODE/IM} and $ M = \beta^{-2}-1$, one can obtain
\begin{align}
\begin{split}
    Q_{6}=2^{\frac{10}{h}+1}J_{5,6} \frac{h^5}{\beta^6} \bigg[\frac{1}{2}\sigma_{3}-\frac{h-5}{2h}\bigg(&\sigma_{1}\sigma_{2}-\frac{2h-5}{6h}\sigma_{1}^{3}+\frac{5(3 h-10)}{48 h}\sigma_{1}^{2}\bigg)\\
    &+\frac{5}{16h}(h-6)\sigma_{2}+\frac{1}{2}c_{6}^{(1)}\sigma_{1}+\frac{1}{2}c_{6}^{(2)}\bigg].
\end{split}
\end{align}
If we set rank $r=4$ and apply the Miura transformation \eqref{eq: D4_Miura}, one can obtain the correspondence
\begin{equation*}
    Q_{6}=2^{\frac{10}{h}+1}J_{5,6}\frac{h^5}{\beta^6}\;I_{6}.
\end{equation*}
We can also predict that $I_{6}$ with general rank is given by
\begin{equation}
    I_{6}=\frac{1}{2}\left[\sigma_{3}-\frac{h-5}{h}\bigg(\sigma_{1}\sigma_{2}-\frac{2h-5}{6h}\sigma_{1}^{3}+\frac{5(3 h-10)}{48 h}\sigma_{1}^{2}\bigg)+\frac{5}{8h}(h-6)\sigma_{2}+c_{6}^{(1)}\sigma_{1}+c_{6}^{(2)}\right].
\end{equation}
Based on the above observations, the ODE/IM correspondence between the WKB integrals and the IoMs is expressed as
\begin{equation}
    Q_{k}= (-1)^{\frac{k}{2}-1}2^{\frac{2k-2}{h}+1}\frac{h^{k-1}}{\beta^{k}}J_{k-1, k}\;I_{k}.
\end{equation}

Finally, there is an extra WKB integral $Q_{r}'$ in Eq.\eqref{eq: Dr_Q}. It leads to the spin-$r$ IoM by the current $R_r$ \eqref{eq: extra_gen}. The correspondence is expressed as
\begin{align}
    Q'_r&=2 J(-\frac{1}{2}, -r)\frac{h^{r}}{\beta^{r}} \; \tilde{\Delta}_r.
\end{align}

Above all, we have obtained the relation between the IoMs and the WKB integrals for $WD_r$ algebras, which confirms the ODE/IM correspondence for the ODE with the pseudo-differential operator.

\section{Conclusions and Discussion}
 In this paper, we explored the WKB solution to the linear problem for the $\hat{\mathfrak{g}}$ affine Toda field equations, especially $\hat{\mathfrak{g}}=A_r^{(1)}$ and $D_r^{(1)}$, modified by the conformal transformation. 
 After taking the conformal and light-cone limit, the linear problem reduces to the higher-order ODE which includes the pseudo-differential operator for $D_r^{(1)}$ case. The WKB expansions of the solutions are known to be the classical conserved currents that appear in the Drinfeld-Sokolov reduction of the adjoint ODE. We computed the period integral of those WKB expansions around the Pochhammer contour systematically up to the $8\text{th}$ order for $A_r^{(1)}$ and $6\text{th}$ order for $D_r^{(1)}$.
From the ODE/IM correspondence, these integrals are expected to agree with the eigenvalue of the quantum IoMs on the vacuum state in the $WA_r$ and $WD_r$ algebras up to some normalization factors. We have constructed the quantum IoMs for these CFTs up to the $5\text{th}$ level and find that they agree with the WKB integrals, which provides strong evidence for the ODE/IM correspondence.

It is interesting to explore the correspondence between the linear problem associated with the affine Toda field equation based on $\hat{\mathfrak g}$ and the quantum IoMs for general $W\hat{\mathfrak g}^{\vee}$-algebra\cite{Ito:1995ny,Keller:2011ek}. So far, we have only computed the eigenvalue of the quantum IoMs on the vacuum state, it is also worth considering the excited states \cite{Ashok:2024zmw}. It is also interesting to apply the supersymmetric affine Toda field theory \cite{Ito:2022cev} and modify the potential term $p(z)$ to extend the dictionary of the ODE/IM correspondence. Finally, we also believe the explicit construction of the IoMs in $WA_{r}$ and $WD_{r}$ CFTs will be valuable for the analysis of thermal correlators in CFT with $W$-symmetry.

\subsection*{Acknowledgments}
The authors would like to thank S.~L.~Lukyanov for constructive criticism and useful explanations for the period integral around the Pochhammer contour. 
The authors are also grateful to Yasuyuki Hatsuda for the useful discussion.
The work of K.I. is supported in part by Grant-in-Aid for Scientific Research 21K03570 from Japan Society for the Promotion of Science (JSPS).

\appendix

\section{Higher-order WKB integrals}\label{app:S_fun}
In this appendix, we present the results of higher-order WKB integrals $Q_{k}$ with $k=6,7,8$ for the ODE associated with $A_{r}^{(1)}$ algebras and $Q_6$ for $D_{r}^{(1)}$ algebras.

\subsection{Higher-order WKB integrals in $A_{r}^{(1)}$-type ODEs}
In Eq.\eqref{eq: Ar_L}, we have shown the formula for $L_{i}$ up to $i=5$ from the $A_r^{(1)}$-type ODE. Here we present  the formula for $L_6$, $L_7$ and $L_8$:
\begin{align}
    \begin{split}
        L_{6} & = s_{6}+\frac{5}{2}(r-4)s_{5}+\frac{1}{24}(r-4)(r-3) (r+70)s_4\\
        & + \frac{5}{48} (r-4) (r-3) (r-2) (r+20)s_{3}\\
        & + \frac{1}{5760}(r-4) (r-3) (r-2) (r-1) (5r^2+698r+5760)s_{2}\\
        & + \frac{1}{2903040}(r-4) (r-3) (r-2) (r-1) r (r+1) (r+2) (35 r^2+7238 r+103560),
    \end{split}
\end{align}
\begin{align}\label{eq: Ar_L_78}
    \begin{split}
        L_{7} & = -s_{7}-3(r-5)s_{6}-\frac{1}{24} (r-5) (r-4) (r+102)s_{5}\\
        & -\frac{1}{8} (r-5) (r-4) (r-3) (r+30)s_4\\
        & - \frac{1}{5760} (r-5)(r-4) (r-3) (r-2) (5r^2+1080r+13152)s_{3}\\
        & - \frac{1}{1920}(r-5)(r-4) (r-3) (r-2) (r-1) (5r^2+1490r+1920)s_{2}\\
        & - \frac{1}{967680}(r-5)(r-4) (r-3) (r-2) (r-1) r (r+1) (r+2) (35 r^2+3038 r+33000),
    \end{split}
\end{align}
\begin{align}
    \begin{split}
        L_{8} & = s_{8}+\frac{7}{2}(r-6)s_{7}+\frac{1}{24} (r-6) (r-5) (r+140)s_{6}\\
        & +\frac{7}{48} (r-6) (r-5) (r-4) (r+42)s_{5}\\
        & + \frac{1}{5760} (r-6)(r-5) (r-4) (r-3) (5 r^2+1398 r+25984)s_{4}\\
        & + \frac{7}{11520}(r-6)(r-5) (r-4) (r-3) (r-2) (5r^2+418r+4032)s_{3}\\
        & + \frac{1}{2903040}(r-6)(r-5) (r-4) (r-3)(r-2) (r-1)\\
        & \times (35 r^3+14658 r^2+539800 r+2903040)s_{2}\\
        & + \frac{1}{1393459200}(r-6)(r-5)(r-4) (r-3) (r-2) (r-1) r (r+1) (r+2)\\
        & \times (175 r^3+97230 r^2+5144984 r+45632832).
    \end{split}
\end{align}

In Eq.\eqref{eq: Ar_Q}, we have shown the WKB integral up to the $5\text{th}$ order. Here we put the $6\text{th}$, $7\text{th}$ and the $8\text{th}$ WKB integrals, which predict the corresponding vacuum eigenvalues of the IoMs:
\begin{align*}
\begin{split}
        Q_{6}&=\frac{1}{h}J_{5,6}\Big(L_{6}+(h-5) \Big[-\frac{1}{2 h}L_{3}^{2}- \frac{1}{h}L_{2}L_{4}+\frac{2h-5}{6h^2}L_{2}^{3} +\frac{5}{2}L_{5}\\
        &\quad-\frac{1}{2h}(5h-13)L_{2}L_{3}-\frac{5}{24}(3hM-8h+41)L_4\\
        &\quad + \frac{ 1}{48 h}(15 h^2 M-40 h^2-35 h M+205 h-257)L_2^2 -\frac{15}{16}(h-3)(h M+3) L_3\\
        &\quad - \frac{1}{1152}(10 h^4 M^3+10 h^4 M^2-22 h^3 M^3-113 h^3 M^2+288 h^3 M+192 h^3\\
        &\quad +205 h^2 M^2-1750 h^2 M-730 h^2+2410 h M-235 h+1895)L_{2}\\
        &\quad + \frac{1}{580608}(h-1) h (h M-5) (h M-1)(8 h^5 M^3+8 h^5 M^2+48 h^4 M^3\\
        &\quad +96 h^4 M^2+48 h^4 M-152 h^3 M^3-474 h^3 M^2-474 h^3 M-152 h^3+478 h^2 M^2\\
        &\quad +969 h^2 M+478 h^2-309 h M-309 h+61)\Big]\Big),
\end{split}
\end{align*}
\begin{align}\label{eq: Ar_Q7}
\begin{split}
        Q_{7}&=\frac{1}{h}J_{6,7}\Big(-L_{7}+(h-6) \Big[\frac{1}{h}L_{3}L_{4}+\frac{1}{h}L_{2}L_{5}- \frac{h-3}{h^2}L_{2}^{2}L_{3}-3L_{6}+\frac{3 (h-3)}{2 h}L_3^2\\
        &\quad- \frac{(h-3)(h-2)}{h^2}L_{2}^{3}+\frac{1}{2}((2 M-5)h+33)L_5-\frac{(h-3) ((2 M-5)h+18)}{2 h}L_2 L_3\\
        &\quad +\frac{(3 h-10)}{h}L_2 L_4+2 (h-4)(h M+4)L_4-\frac{(h-3) (h-2) (h M+2)}{h}L_2^2\\
        &\quad +\frac{1}{40} (h-3) \left(h^3 M^3+h^3 M^2-12 h^2 M^2+31 h^2 M+20 h^2-195 h M-42 h-254\right)L_{3}\\
        &\quad +\frac{1}{40} (h-3) (h-2)(h M+2) \left(h^2 M^2+h^2 M-14 h M-11 h-7\right)L_2\Big]\Big).
\end{split}
\end{align}
In Sec.\ref{Sec: 4}, we have given the prediction for $i_{6}$ in terms of the momentum parameter $\sigma_{i}$. It is also possible to rewrite $i_{6}$ in terms of $\Delta_{i}$.
\begin{equation*}
   i_{6}= \Delta_{6} + a_{6}^{(1)}\left\{\Delta_{4}\Delta_{2}+\frac{1}{2}\left[\Delta_{3}^{2}+a_{6}^{(2)}\left(\Delta_{2}^{3}+a_{6}^{(3)}\Delta_{2}^{2}+a_{6}^{(4)}\Delta_{2}+a_{6}^{(5)}\right) \right]  \right\} 
\end{equation*}
with the tedious coefficients
\begin{align*}
    a_{6}^{(2)}&= \frac{(-101 + 5 h + 29 h^2 - 5 h^3)c+2 (-1 + h) (-37 - 48 h + 25 h^2)}{3(5 c+22) h \left(h^2-1\right)},\\
    a_{6}^{(3)}&= \frac{(-5 h^3+29 h^2+5 h-101)c^{2}+(5 h^4-8 h^3+69 h^2+104 h-626)c}{8 \left(5 h^3-29 h^2-5 h+101\right)c-16 (-1 + h) (-37 - 48 h + 25 h^2)}\\
    &-\frac{2 (h-1) \left(h^3-157 h^2+282 h+188\right)}{8 \left(5 h^3-29 h^2-5 h+101\right)c-16 (-1 + h) (-37 - 48 h + 25 h^2)},\\
    a_{6}^{(4)}&= \frac{5 \left(5 h^3-29 h^2-5 h+101\right)c^{3}}{960 (101 - 5 h - 29 h^2 + 5 h^3)c-1920 (h-1) \left(25 h^2-48 h-37\right)}\\
    &-\frac{2 \left(25 h^4-165 h^3+698 h^2+650 h-3662\right)c^{2}}{960 (101 - 5 h - 29 h^2 + 5 h^3)c-1920 (h-1) \left(25 h^2-48 h-37\right)}\\
    &+\frac{(25 h^5-295 h^4-1843 h^3+6103 h^2-6418 h+10948)c}{960 (101 - 5 h - 29 h^2 + 5 h^3)c-1920 (h-1) \left(25 h^2-48 h-37\right)}\\
    &+\frac{2 (h-1) \left(55 h^4-398 h^3-2147 h^2+5922 h+1248\right)}{960 (101 - 5 h - 29 h^2 + 5 h^3)c-1920 (h-1) \left(25 h^2-48 h-37\right)},
\end{align*}
\begin{align*}
    a_{6}^{(5)}&= \frac{35  \left(5 h^3-29 h^2-5 h+101\right)c^{4}}{483840  \left(5 h^3-29 h^2-5 h+101\right)c-967680 (h-1) \left(25 h^2-48 h-37\right)}\\
    &-\frac{(525 h^5-3815 h^4+12621 h^3+34311 h^2-67134 h-89284)c^{3}}{483840 (h+1) \left(5 h^3-29 h^2-5 h+101\right)c-967680 (h^2-1) \left(25 h^2-48 h-37\right)}\\
    &+\frac{(525 h^6-5880 h^5-12432 h^4+30930 h^3-87505 h^2+154110 h+318892)c^{2}}{483840 (h+1) \left(5 h^3-29 h^2-5 h+101\right)c-967680 (h^2-1) \left(25 h^2-48 h-37\right)}\\
    &-\frac{(h-1) \left(175 h^6-3500 h^5+14388 h^4+133250 h^3-189355 h^2-369510 h-13128\right)c}{483840 (h+1) \left(5 h^3-29 h^2-5 h+101\right)c-967680 (h^2-1) \left(25 h^2-48 h-37\right)}\\
    &-\frac{22 (h-4) (h-3) (h-2) (h-1)^2 \left(35 h^2+112 h+93\right)}{483840 (h+1) \left(5 h^3-29 h^2-5 h+101\right)c-967680 (h^2-1) \left(25 h^2-48 h-37\right)}.
\end{align*}
Here $a_{6}^{(1)}$ is absent due to the difficulty in constructing $\Delta_{6}$. But when $r=1$, $2$, and $3$, it leads to the result in Eq.
\eqref{r1_i6}, Eq.\eqref{r2_i6} and Eq.\eqref{r3_i6}  after some normalization.

Next, we turn to $Q_{7}$ and $Q_{8}$. Based on Eqs.\eqref{eq: Ar_L_78} and \eqref{eq: Ar_Q7}, one can obtain the WKB integral in terms of momentum parameter $s_{i}$
\begin{align*}
    Q_7 = J_{6,7}&\Big[ \frac{1}{h}s_{7}+(h-6)\Big(-\frac{1}{h^2}(\sigma_{2}\sigma_{5}+\sigma_{3}\sigma_{4})+\frac{ h-3}{h^3}\sigma_{2}^{2}\sigma_{3}+\frac{ h-3}{h}(1+M)\sigma_{2}\sigma_{3}\\
    &\quad-(1+M)\sigma_{5}-\frac{h(h-3) (1+M)\left(hM^2(1+M) -12\right)}{40 }\sigma_{3} \Big) \Big].
\end{align*}
After the plunge of the ODE/IM correspondence \eqref{eq: ODE/IM}, one can predict that the vacuum eigenvalue of the IoM $i_{7}$ is of the form
\begin{align*}
    i_7 =  \sigma_{7}-(h-6)\Big(\frac{1}{h}(\sigma_{2}\sigma_{5}+\sigma_{3}\sigma_{4})-\frac{ h-3}{h^2}(\sigma_{2}^{2}\sigma_{3}+\sigma_{2}\sigma_{3})+\frac{1}{h}\sigma_{5}+\frac{  (h-3) \left(h\alpha_{0}^2-12\right)}{40 h^2}\sigma_{3} \Big).
\end{align*}

Finally, we can obtain $Q_{8}$ after following a similar step with $Q_{7}$.
\begin{align*}
    Q_{8}=&J_{7,8}\Big[\frac{1}{h}s_{8}+(h-7)\Big(-\frac{1}{h^2}\left(s_{6}s_{2}+s_{5}s_{3}+\frac{1}{2}s_{4}^2-\frac{2h-7}{2h}s_{2}^2 s_{4}+\frac{(2 h-7) (3 h-7)}{24h^2}s_{2}^4\right)\\    
    &\quad -\frac{35}{24}(1+M)s_{6}+\frac{2 h-7}{2 h}(1+M)s_{2}s_{3}^2-\frac{7(2 h-7) (5 h-11)}{144 h^2}(1+M)s_{2}^3\\
    &\quad +\frac{7 (5 h-19)}{24 h}(1+M)s_{2}s_{4}\\
    &\quad -\frac{7h (1+M)\left(14 h^2 M^2-54 h M^2-211 h M-211 h+791 M+791\right)}{1920 }s_{4}\\
    &\quad +\frac{7(1+M)}{11520 }(42 h^3 M^2-232 h^2 M^2-633 h^2 M-633 h^2   \\
    &\quad +302 h M^2 +3524 h M+3524 h-4627 M-4627)s_{2}^2 \\
    &\quad +\frac{h^2(1+M)}{414720}(56 h^5 M^4+392 h^4 M^4-3416 h^3 M^4-9982 h^3 M^3-9982 h^3 M^2   \\
    &\quad +4888 h^2 M^4+54624 h^2 M^3+96361 h^2 M^2+83474 h^2 M+41737 h^2\\
    &\quad -70322 h M^3-299978 h M^2-459312 h M-229656 h \\
    &\quad +297479 M^2+594958 M+297479 )s_{2}\\
    &\quad +\frac{h^4 (h-1)(1+M)}{199065600}(96 h^7 M^6+768 h^6 M^6-1584 h^5 M^6-8176 h^5 M^5\\
    &\quad -8176 h^5 M^4-18048 h^4 M^6-27244 h^4 M^5-27244 h^4 M^4+36048 h^3 M^6\\
    &\quad +343504 h^3 M^5+695436 h^3 M^4+703864 h^3 M^3+351932 h^3 M^2-526484 h^2 M^5\\
    &\quad -2474612 h^2 M^4-4512137 h^2 M^3-3795771 h^2 M^2-1847643 h^2 M-615881 h^2\\
    &\quad +2540356 h M^4+8489936 h M^3+12768028 h M^2+10227672 h M\\
    &\quad +3409224 h-4445623 M^3-13336869 M^2-13336869 M-4445623)  \Big)  \Big].
\end{align*}
Substituting the relation \eqref{eq: ODE/IM}, one can predict the vacuum eigenvalue of the IoM $I_{8}$ as
\begin{align*}
    I_{8}=& \sigma_{8}+(h-7)\Big(-\frac{1}{h}\left(\sigma_{6}\sigma_{2}+\sigma_{5}\sigma_{3}+\frac{1}{2}\sigma_{4}^2-\frac{2h-7}{2h}\sigma_{2}^2 \sigma_{4}+\frac{(2 h-7) (3 h-7)}{24h^2}\sigma_{2}^4\right)\\    
    &\quad -\frac{35}{24h}\sigma_{6}+\frac{2 h-7}{2 h^2}\sigma_{2}\sigma_{3}^2-\frac{7(2 h-7) (5 h-11)}{144 h^3}\sigma_{2}^3+\frac{7 (5 h-19)}{24 h^2}\sigma_{2}\sigma_{4}\\
    &\quad -\frac{7 \left(h \left(2(7 h-27)\alpha_{0}^2-211\right)+791\right)}{1920 h^2}\sigma_{4}\\
    &\quad +\frac{7 \left(h \left(2 \alpha_0^2 (h (21 h-116)+151)-633 h+3524\right)-4627\right)}{11520 h^3}\sigma_{2}^2 \\
    &\quad -\frac{1}{414720 h^3}\big(56 \alpha_0^4 h^5+392 \alpha_0^4 h^4-14 \alpha_0^2 \left(244 \alpha_0^2+713\right) h^3 \\
    &\quad -14 \left(5023 \alpha_0^2+16404\right) h+\left(4888 \alpha_0^4+54624 \alpha_0^2+41737\right) h^2+297479\big)\sigma_{2} \\
    &\quad +\frac{h-1}{199065600 h^3}\big( 96 \alpha _0^6 h^7+768 \alpha _0^6 h^6-1584 \alpha _0^6 h^5-8176 \alpha _0^4 h^5-18048 \alpha _0^6 h^4 \\
    &\quad -27244 \alpha _0^4 h^4+36048 \alpha _0^6 h^3+343504 \alpha _0^4 h^3+351932 \alpha _0^2 h^3-526484 \alpha _0^4 h^2\\
    &\quad -1948128 \alpha _0^2 h^2-615881 h^2+2540356 \alpha _0^2 h+3409224 h-4445623 \big)\Big).
\end{align*}
The result agrees with the one in \cite{Ashok:2024zmw} when $h=3$. Further $Q_k$ can be calculated similarly.

\subsection{Higher-order WKB integrals in $D_{r}^{(1)}$-type pseudo ODEs}
Finally, we present $Q_{6}$ in terms of $L'_{i}$ in $D_{r}^{(1)}$-type pseudo ODEs
\footnotesize
\begin{align*}
        Q_{6} & = -\frac{2^{-\frac{10}{h}}}{h}J_{5,6}\Big(L'_{6}+(h-5) \Big[-\frac{1}{2 h}L_{3}^{'2}- \frac{1}{h}L'_{2}L'_{4}+\frac{2h-5}{6h^2}L_{2}^{'3}\Big]\\
        &\quad +\frac{5}{2}(h-4)L'_{5}-\frac{1}{2h}(h-5)(5h-13)L'_{2}L'_{3}\\
        &\quad -\frac{5}{24}(3 hM (h-6)-8 h^2+57 h-118)L'_4\\
        &\quad + \frac{ h-5}{48 h}(15 h^2 M-40 h^2-50 h M+85 h-62)L_2^{'2} \\
        &\quad -\frac{15}{16}(h-2)(3 hM (h-6)+h-22) L'_3\\
        &\quad - \frac{1}{1152}(10 h^5 M^3+10 h^5 M^2-72 h^4 M^3-163 h^4 M^2+200 h^3 M^3++980 h^3 M^2\\
        &\quad +288 h^4 M-2470 h^3 M-2300 h^2 M^2+3960 h^2 M-3320 h M\\
        &\quad +192 h^4-1690 h^3+2695 h^2-740 h-820)L'_{2}\\
        &\quad + \frac{1}{580608}h (h+2) (h M-5) (h M-1)(8 h^6 M^3-16 h^5 M^3-368 h^4 M^3\\
        &\quad +8 h^6 M^2+32 h^5 M^2+48 h^5 M-1074 h^4 M^2+1888 h^3 M^3+5320 h^3 M^2\\
        &\quad -5672 h^2 M^2-858 h^4 M+5139 h^3 M-11652 h^2 M+3228 h M\\
        &\quad -152 h^4+1694 h^3-5939 h^2+3268 h-572)\Big).
\end{align*}
\normalsize
The result in terms of $s_{i}$ and the correspondence analysis has been given in \eqref{eq: Dr_Q6}.

\section{Construction of primary $W$-currents}
In this Appendix, we will explain an explicit construction of the primary W-currents.
In general, the spin $k$ current $\Tilde{W}_k(z)$ obtained from the Miura transformation is not a primary field. 
But we can redefine these fields into primary fields by adding lower spin fields and their derivatives. For $\Tilde{W}_k(z)$ the primary field is  denoted by $W_k(z)$. Here we present our results in $WA_{r}$ and $WD_{r}$ algebras with lower ranks.
\subsection{Primary $W$-currents in $WA_r$ algebras}\label{app:pwa}
Computing the OPEs in lower rank $WA_r$ algebras\footnote{We used Mathematica package\cite{Thielemans:1991uw} for computation of OPE.}, it is possible to infer the primary $W$-currents:
\begin{align}\label{eq: primary_Miura}
    \begin{split}
        &T(z)=\Tilde{W}_2(z),\\
        &W_{3}(z)=\Tilde{W}_{3}(z)-\frac{h-2}{2}\alpha_{0}\partial_{z}T(z),\\
        &W_{4}(z)=\Tilde{W}_{4}(z)+(h-3)\left(x_{1}\altcolon TT\altcolon(z)-\frac{1}{2}\alpha_{0}\partial_{z}\Tilde{W}_{3}(z)-x_{2}\partial_{z}^{2}T(z)\right),\\
        &W_5(z)=\tilde{W}_5(z)-{r-3\over2}\alpha_{0}\partial_{z} \tilde{W}_4(z)+y_1 \altcolon T W_3\altcolon(z)+y_2 \partial_{z}^2 W_3(z)+y_3\alpha_{0} \partial_{z}^3 T(z),
    \end{split}
\end{align}
with 
\begin{align}
    \begin{split}
        &x_{1}=\frac{(h-2) ((5h+7)c-2h+2)}{2(5c+22)h \left(h+1\right)\left(h-1\right)},\quad x_{2}=\frac{(h-2) (2c^{2}+(h+15)c-10 (h-1))}{4 (5 c+22) h \left(h+1\right)\left(h-1\right)},\\
        &y_1=\frac{(h-4) (h-3) ( (7 h+13)c-6 h+6)}{(7 c+114) h \left(h+1\right)\left(h-1\right)},\\
        &y_2=-\frac{3 (h-4) (h-3) (c^2+ (h+21)c-18 (h-1))}{4 (7 c+114) h \left(h+1\right)\left(h-1\right)},\\
        &y_3=\frac{(h-4) (h-3) \left((32-7 h)c^2+( 7 h^2-117h+620)c+6 (h-1) (19 h-92)\right)}{24 (7 c+114) h \left(h+1\right)\left(h-1\right)}.
    \end{split}
\end{align}
Correspondingly, the relations between conformal weights $\Delta_{i}$ of primary $W_k(z)$ and $\Tilde{\Delta}_{i}$ can be derived from Eq.\eqref{eq: primary_Miura}.
    \begin{align}\label{eq: Delta_Primary_Miura}
    \begin{split}
        \Delta_{2} &= \Tilde{\Delta}_{2},\\
        \Delta_{3} &= \Tilde{\Delta}_{3}+(h-2)\alpha_{0}\Delta_{2},\\
        \Delta_{4} &= \Tilde{\Delta}_{4}+(h-3)\left(c_{4}^{(1)}\Delta_{2}^{2}+\frac{3}{2}\alpha_{0}\Tilde{\Delta}_{3}+c_{4}^{(2)}\Delta_{2}\right),\\
        \Delta_{5} &= \Tilde{\Delta}_{5} + (h-4)\left( c_{5}^{(1)}\Delta_{2}\Delta_{3} + 2\alpha_{0}\Tilde{\Delta}_{4} + c_{5}^{(2)}\Delta_{3}-(h-3)(h-2)\alpha_{0}^{2}\Delta_{2} \right)
    \end{split}
    \end{align}
with the coefficients
\begin{align*}
    \begin{split}
        &c_{4}^{(1)}=\frac{(h-2)[(5h+7)c-2h+2]}{2 (5c+22) h \left(h+1\right)\left(h-1\right)}, \quad c_{4}^{(2)}= -\frac{(h-2)[6c^{2}-(7 h-31)c-26 (h-1)]}{2 (5 c+22) h \left(h+1\right)\left(h-1\right)},\\
        &c_{5}^{(1)}=\frac{(h-3) [(7 h+13)c-6 h+6]}{(7 c+114) h \left(h+1\right)\left(h-1\right)},\quad c_{5}^{(2)}= -\frac{3 (h-3) \left[3 c^2-4 (c+12) h+50 c+48\right]}{(7 c+114) h \left(h+1\right)\left(h-1\right)}.
    \end{split}
\end{align*}

\subsection{Primary $W$-currents in $WD_r$ algebra}\label{app:wdr}
Here we first discuss a recursive construction of $W$-currents in $WD_r$ algebra.
We first compute the OPE of $V_r^{(r)}(z)$ using the relation
\begin{align}
    R_r^{(r)}(z)V_r^{(r)}(w)&=(-\tilde{p}_r R^{(r-1)}_{r-1}+\alpha_0\partial R_{r-1}^{(r-1)})(z)
    (-\tilde{p}_r R^{(r-1)}_{r-1}+\alpha_0\partial R_{r-1}^{(r-1)})(w).
\end{align}
We then use the formula \eqref{eq:opev1} for $r-1$ in the RHS of the OPE, which leads to the recursion relations for the constant $A_r$ and the W-currents $W_{2k}^{(r)}$:
\begin{align}
A_{r}&=A_{r-1}-(2r-1)(2r-2)a^2 A_{r-1},
\end{align}
\begin{align}
W_{2}^{(r)}&=W^{(r-1)}_2+{1\over2}(\tilde{p}_r\tilde{p}_r)-(r-1)a\partial \tilde{p}_r,
\end{align}
\begin{align}
W_4^{(r)}&=W_4^{(r-1)}+{1\over4}(1-(2r-3)(2r-4)a^2)\Bigl\{(\partial^2\tilde{p}_r\tilde{p}_r)-{1\over2}\partial^2(\tilde{p}_r \tilde{p}_r)\Bigr\}
\nonumber\\
&+(\tilde{p}_r(\tilde{p}_r W^{(r-1)}_2))
-2(r-2)a (\partial \tilde{p}_r W_2^{(r-1)})-a (\tilde{p}_r \partial W_2^{(r-1)})
\nonumber\\
&+{1\over12}(r-1) a (1-(2r-3)(2r-4)a^2)\partial^3 \tilde{p}_r +(r-2)a^2\partial^2 W_2^{(r-1)}.
\end{align}
For $r=1$, we define 
\begin{align}
W_2^{(1)}&={1\over2}(\tilde{p}_1\tilde{p}_1),\\
W_4^{(1)}&={1\over4}(1-2a^2) \Bigl\{(\partial^2\tilde{p}_1\tilde{p}_1)-{1\over2}\partial^2(\tilde{p}_1\tilde{p}_1)\Bigr\}.
\end{align}
Thus we obtain the non-primary $W$-currents $\Tilde{W}_4$. 
The primary $W_{4}$ current is given by
\begin{align}
\begin{split}
W_4(z)&=\Tilde{W}_4(z)+a_1 \altcolon TT \altcolon(z) +a_2 \partial_{z}^2 T(z)
\end{split}
\end{align}
with the coefficients
\begin{align}
a_1&=-{5r+1\over 22+5c}\left( 1 -2(r-2)(2r-3)\alpha_{0}^2\right),\nonumber \\
a_2&={1\over 22+5c}\left( 1 -2(r-2)(2r-3)\alpha_0^2\right)
\left({4r+5\over2}-r(r-1)(2r-1)\alpha_0^2\right).
\label{eq:coeffa1a2}
\end{align}
The relation between $\Delta_{4}$ of the primary $W_4(z)$ and $\Tilde{\Delta}_{4}$ is expressed as
\begin{equation}
    \Delta_{4} = \Tilde{\Delta}_{4} + a_{1}\Delta_{2}^{2}+ 2(a_{1}+3a_{2})\Delta_{2},
\end{equation}
where $a_1$ and $a_2$ is given in \eqref{eq:coeffa1a2}.

For $W_{6}(z)$, so far, we can only give its value in $WD_{4}$ algebra. The primary spin 6 current $W_6$ is given by
\begin{align}\label{eq: D4_Miura}
    \begin{split}
        W_6&=\tilde{W}_{6}-{1\over3} \altcolon T \tilde{W}_4\altcolon-{ -13 + 1388 \alpha_{0}^2 - 22560 \alpha_{0}^4\over 84 (4 - 241 \alpha_{0}^2 + 2352 \alpha_{0}^4)} \altcolon T(\altcolon TT\altcolon) \altcolon\\
&-{9 - 2992 \alpha_{0}^2 + 70688 \alpha_{0}^4 - 288960 \alpha_{0}^6\over 
   168 (4 - 241 \alpha_{0}^2 + 2352 \alpha_{0}^4)} \altcolon T\partial_{z}^2 T\altcolon\\
&-{23 - 815 \alpha_{0}^2 + 6302 \alpha_{0}^4 + 15960 \alpha_{0}^6\over 
   84 (4 - 241 \alpha_{0}^2 + 2352 \alpha_{0}^4)}\altcolon \partial T \partial T\altcolon
  \\
   &+{ 1 - 4 \alpha_{0}^2\over 6} \partial^2 \tilde{W}_4-{-2 + 425 \alpha_{0}^2 - 10274 \alpha_{0}^4 + 53496 \alpha_{0}^6 - 40320 \alpha_{0}^8\over 
   144 (4 - 241 \alpha_{0}^2 + 2352 \alpha_{0}^4)} \partial^4 T,\\
   \Delta_6&=\tilde{\Delta}_6-{1\over3}\Delta_2\tilde{\Delta}_4+t_3\Delta_2^3+t_4\tilde{\Delta}_4+t_5\Delta_2^2+t_6\Delta_2,
    \end{split}
\end{align}
with
\begin{align*}
t_3&={13 - 1388 \alpha_{0}^2 + 22560 \alpha_{0}^4\over 84 (4 - 241 \alpha_{0}^2 + 2352 \alpha_{0}^4)},\\
t_4&=2-{40\alpha_{0}^3\over3},\\
t_5&={-41 + 3908 \alpha_{0}^2 - 101912 \alpha_{0}^4 + 
 803040 \alpha_{0}^6\over 84 (4 - 241 \alpha_{0}^2 + 2352 \alpha_{0}^4)},\\
 t_6&={-1 - 30 \alpha_{0}^2 + 6796 \alpha_{0}^4 - 186480 \alpha_{0}^6 + 
 1411200 \alpha_{0}^8\over 42 (4 - 241 \alpha_{0}^2 + 2352 \alpha_{0}^4)}.
\end{align*}
The expansion of $\Tilde{\Delta}_{6}$ in terms of momentum becomes
\begin{align}
\Delta_6&={1\over2}\sigma_{3}+(-1+{11\over2}\alpha_{0}^2) \sigma_{2}
+({1\over2}-8a^2+{61\over2}\alpha_{0}^4)\sigma_{1}-7 \alpha_{0}^2 + 161 \alpha_{0}^4 - {1429 \alpha_{0}^6 \over2}.
\nonumber
\end{align}

\section{Normal orderings on the cylinder}\label{app:ope1}
In this Appendix,  we show explicitly how we obtain some important formulas for normal orderings and their zero modes on the cylinder from Eq.\eqref{eq:opeab1}.
After introducing the Bernoulli polynomial $\psi_n(x)$ of the second kind by
\begin{align}
\frac{(z+1)^x }{\log(1+z)}
&=\sum_{n=0}^{\infty}\psi_n(x)z^{n-1}, \quad |z|<1
\end{align}
and the OPE
\begin{align}
  A_R(z)B_R(w)&=\sum_{k=0}^{\Delta_A+\Delta_B}  {\{A_RB_R\}_k(w) \over (z-w)^k}
  +\{A_R B_R\}_0(w)+\cdots,
\end{align}
we expands the integrand in Eq.\eqref{eq:opeab1} at $z=w$. Following the procedure in \cite{Novaes:2021vjh}, one obtains
\begin{align}
     :\hat{A}\hat{B}:(v)&=
     \hat{A}_-(v) \hat{B}(v)+\hat{B}(v)\hat{A}_+(v)
     +\sum_{k=1}^{\Delta_A+\Delta_B} f_k(\Delta_A-1) 
     \{A_R B_R\}_k(w) w^{\Delta_A+\Delta_B-k},
     \label{eq:no_on_cyl1}
\end{align}
where
\begin{align}
f_k(x)&=\psi_k(x)-{(x)_k\over k!},\\
    \hat{A}_+(v)&=\sum_{n=0}^{\infty}\hat{A}_n e^{-nv},\quad \hat{A}_-(v)=\sum_{n=1}^{\infty}\hat{A}_{-n} e^{nv},
\end{align}
and $(a)_n=a(a-1)\cdots (a-n+1)$. Some values of $f_k(x)$ are seen in Table \ref{tab:ber1}.
\begin{table}[h]
\begin{center}
\begin{tabular}{c|cccccc}
$k$ & 1& 2& 3& 4& 5 & 6\\\hline
$f_k(1)$ & ${1\over2}$ & ${5\over12}$ & $-{1\over24}$ & ${11\over720}$ & $-{11\over 1440}$ & ${271\over 60480}$\\ 
$f_k(2)$ & ${1\over2}$ & ${11\over12}$ & ${3\over8}$ & $-{19\over720}$ & ${11\over 1440}$ & $-{191\over 60480}$\\
\end{tabular}
\caption{The values of $f_k(x)=\psi_k(x)-{(x)_k\over k!}$ for $x=1$ and $2$.}
\end{center}
\label{tab:ber1}
\end{table}

It is convenient to rewrite the first two terms in \eqref{eq:no_on_cyl1} in terms of the normal ordered product on the complex plane to compute multiple normal ordered products.
It is found that
\begin{align}
\begin{split}
    \hat{A}_-(v) \hat{B}(v)+\hat{B}(v)\hat{A}_+(v)
    &=w^{\Delta_A+\Delta_B}\{A_R B_R\}_0(w)\\
    &+\sum_{n=1}^{h_A-1}\sum_{k=1}^{\Delta_A+\Delta_B} 
    w^{\Delta_A+\Delta_B-k} {(\Delta_A-n-1)_{k-1} \over (k-1)!}
    \{A_R B_R\}_k(w).
    \label{eq:no_on_cyl2}
\end{split}
\end{align}
Based on the technique above, let us show some important normal ordering in the local IoMs. The first one is 
\begin{align}
    :\hat{T}\hat{T}:(v)&=w^4 \{TT\}_0(w)-{c-10\over 12} w^2 T(w)
    +{3\over2}w^3 \partial T(w) +{22c+5c^2\over2880}.
\end{align}
Its zero mode is given by
\begin{align}
    (:\hat{T}\hat{T}:)_0&=2\sum_{n=1}^{\infty}L_{-n}L_n +L_0^2-{c+2\over12}L_0+{5c^2+22c \over 2880}.
    \label{eq:3-68}
\end{align}
The formula above determines the normal ordered product of the energy-momentum tensors on the cylinder in terms of that on the complex plane.
One can further calculate the normal ordered product of $\hat{T}$ and $:\hat{T}\hat{T}:$ from the OPE of $T_R(z)(\altcolon TT\altcolon)_R(w)$ as
\begin{align}
:\hat{T}(:\hat{T}\hat{T}:):(v)&=\hat{T}_- :\hat{T}\hat{T}:(v)
+:\hat{T}\hat{T}:(v) \hat{T}_+
+\sum_{k=1}^{6}f_k(1)\{T_R (\altcolon TT\altcolon)_R\}_k(w),
\label{eq:3-58}
\end{align}
where $T_R(w)=T(w)-{c\over 24w^2}$, and
\begin{align}
    :\partial \hat{T}\partial \hat{T}:(v)&=
    \partial\hat{T}_- \partial \hat{T}:(v)
+\partial\hat{T}:(v) \partial \hat{T}_+
+\sum_{k=1}^{6}f_k(2)\{\partial T_R \partial T_R\}_k(w).
\label{eq:3-59}
\end{align}
Their zero modes are
\begin{align}
    (:\hat{T}(:\hat{T}\hat{T}:):)_0&=
    \sum_{n=1}^{\infty}\tilde{L}_{-n} (:\hat{T}\hat{T}:)_n +(:\hat{T}\hat{T}:)_{-n}\tilde{L}_n +L_0^3-{c+4\over 8}L_0^2
\nonumber\\
&+\left({c^2\over 192}+{7c \over 160}+{1\over 15}\right)L_0+\left( -{c^3\over 13824}-{11c^2\over 11520}-{47c\over 15120}\right),
\\
    (:\partial \hat{T}\partial \hat{T}:)_0&=-2\sum_{n=1}^{\infty}n^2L_{-n}L_n
+{31c \over 30240}-{1\over 60}L_0.
\end{align}
Next, we give the normal ordered product of $T$ and spin $s$ primary current $W_{s}$.
\begin{align}
    :\hat{T} \hat{W}_s:(v)&=\hat{W}_-(v) \hat{T}(v)+\hat{T}(v)\hat{W}_+(v) +f_1(1) \{ T_R W\}_1(w)+f_2(1)\{T_R W\}_2(w).
\label{eq:3-60}
\end{align}
where $\{T_R W\}_1=\partial W$ and $\{T_R W\}_2=s W$.
Its zero mode of $:\hat{T}\hat{W}:(u)$ is found to be
\begin{align}
     (:\hat{T}\hat{W}_{s}:)_0&=\sum_{n=1}^{\infty}L_{-n}(W_{s})_n+\sum_{n=1}^{\infty}(W_{s})_{-n}L_n
+(W_{s})_0 L_0-{2h+c\over24}(W_{s})_0.
\label{eq:3-69}
\end{align}

Finally, we show the normal ordering $:\hat{W}_3\hat{W}_3:(v)$ which appears in the $6\text{th}$ order IoM.  It takes the form
\begin{align}
    :\hat{W}_3\hat{W}_3:(v)
    &=\hat{W}_{-}(v)\hat{W}(v)+\hat{W}(v)\hat{W}_+(v)
    +\sum_{k=1}^{6}f_k(2)\{W_3 W_3\}_k(w),
\end{align}
The structure of the OPE $W_3(z)W_3(w)$ depends on the $WA_r$-algebra. We refer \cite{Fateev:1987vh} and \cite{Blumenhagen:1990jv,Kausch:1990bn} for low-rank results ($r=2,3$). 
From the normal ordered products of operators on the cylinder, one can calculate the zero modes, we give the results for $WA_2$ and $WA_3$ algebras. For the $WA_2$ algebra, the zero mode is
\begin{align}
    (:\hat{W}_3\hat{W}_3:)_0&=2\sum_{n=-\infty}^{\infty}(W_3)_{-n}(W_3)_n+(W_3)_0^2-{191c\over 181440}+\left({17\over 360}-{b^2\over 30}\right)L_0-{b^2\over 60}L_0^2.
\end{align}
Here $b^2={16\over 22+5c}$. For the $WA_3$ algebra, the zero mode is given by
\begin{align}
    (:\hat{W}_3\hat{W}_3:)_0&=2\sum_{n=-\infty}^{\infty}(W_3)_{-n}(W_3)_n+(W_3)_0^2-{1\over3}(W_4)_0-{191c(c+7)\over 1814400}
    \nonumber\\
    &-{4(c+7) \over 15(22+5c)}(2\sum_{n=1}^{\infty}L_{-n}L_n+L_0^2+2L_0)
+{(c+7)(2102+85c) \over 3600(22+5c)}L_0.
\end{align}
where $L_{n}$ and $(W_{3})_{n}$ are the mode expansion of $T(z)$ and $W_{3}(z)$.


\end{document}